\documentclass[preprint]{aastex}
\revised{10 Feb 2009}
\begin{document}
\title{Luminosity Functions of Type Ia Supernovae and their Host Galaxies
from the Sloan Digital Sky Survey}
\author{Naoki Yasuda$^{1,3}$ and Masataka Fukugita$^{1,2,3}$}
\affil{$^{1}$Institute for Cosmic Ray Research, University of Tokyo, Kashiwa 277-8582, Japan}
\affil{$^{2}$Institute for Advanced Study, Princeton, NJ08540, U.S.A.}
\affil{$^{3}$Institute for the Physics and Mathematics of the Universe, University of Tokyo, Kashiwa 277-8568, Japan}

\begin{abstract}

The sample of 137 low-redshift type Ia supernovae with $0.05\leq z\leq
0.3$ obtained from the SDSS-II Supernova Survey for the southern
equatorial stripe of 300 square degrees is used to derive the
luminosity functions of type Ia supernovae and of their host galaxies
in the $gri$ passbands. We show that the luminosity function of type
Ia supernova host galaxies matches well with that of galaxies in the
general field, suggesting that the occurrence of type Ia supernovae
does not favour a particular type of galaxies but is predominantly
proportional to the luminosity of galaxies. The evidence is weak that
the supernovae rate varies with colour of host galaxies. 
The only evidence that points to possible  correlation between
the supernova rate and star formation activity is that the supernova rate in 
late-type galaxies is higher
than that in early-type galaxies by $31\pm35$\%.  In our low redshift
sample the component of type Ia supernova rate that is proportional to
star formation activity is  
not manifest in the integrated supernova rate, while our observation is
compatible with the current two-component models.  
The sample contains 8
type Ia supernovae whose host galaxies were not identified, but it is
shown that their occurrence is consistent with them occurred in low
luminous galaxies beyond the survey.
The luminosity function of type Ia supernovae is
approximately Gaussian with the full-width half maximum being a factor
of 1.4 in luminosity. The Gaussian distribution becomes tighter if the
ratio of extinction to reddening, $R_V$, is lower than the
characteristic value for the Milky Way and if luminosity is corrected
for the light curve shape. The colour excess is
$\approx0.07$ mag which is
significantly smaller than reddening expected for field galaxies. This
colour excess does not vary with the distance of the supernovae from the
centre of the host galaxy to 15 kpc. This suggests that the major part
of the colour excess appears to be either intrinsic or reddening that
arises in the immediate environment of supernova, rather than
interstellar reddening in host galaxies, and most of type Ia
supernovae take place in relatively dust free environment.

\end{abstract}

\keywords{supernovae : general}

\section{Introduction}

Recent studies have revealed that our understanding of the mechanism
of Type Ia supernovae (SNe Ia) is poorer than we had thought. It is
suspected that there are two different progenitor types
responsible for SNe Ia: explosions from
old systems as in the long-accepted scenario and explosions in 
young stellar systems
\citep{Dallaporta1973, 
Tammann82, Mannucci05, Sullivan06}. The evidence, however, is not
conclusive yet, and some observations at high redshift do not fit this
picture; for example, there are indications for a drop in the SN rate at $z>1.5$,
where star formation rate is still rising 
\citep{Poznanski07, Dahlen08}, which does not support the presence of
a large prompt component.

The present paper studies the luminosity
functions (LF) of SNe Ia and their host galaxies and correlations in
properties of
SN Ia and of host galaxies. \citet{MB90} and
\citet{Richardson02} studied the LF of various types of SNe
based on the Asiago Supernova Catalog \citep{ASC}. In particular,
\citet{Richardson02} showed that the LF of SNe Ia is consistent with
the Gaussian distribution with $\langle M_B\rangle=-19.46+5\log(H_0/60)$ 
and $\sigma = 0.56$
using 111 spectroscopically normal SNe Ia without correcting for the
light curve shape parameter.  With the correction for the decline rate
$\sigma$ can be reduced to 0.11
\citep{Phillips99}.  We are not aware of a study of the LF of SNe Ia
host galaxies. We may hope that a comparison with the LF of field galaxies may
hint us as to what type of galaxies would preferentially host SNe Ia.

We have accumulated a sample of 
SNe Ia acquired in the second phase of
the Sloan Digital Sky Survey \citep[SDSS;][]{York00}. Supernovae
were searched during September to November of 2005 to 2007 
by repeated imaging for the sky area of 300
square degree 
of the southern equatorial region,  $-60\arcdeg<{\rm R.A.}<+60\arcdeg$,
$-1\fdg25<{\rm Decl.}<+1\fdg25$, every two days
\citep{Frieman08,Sako08,Dilday08}.  Approximately
500 SNe Ia at $0.05<z<0.4$ were identified using well-defined selection
criteria and their light curves were measured \citep{Sako08,Holtzman08}.  
The advantage of the use of the SDSS is,
apart from its accurate five colour photometry
\citep{Fukugita96,Smith02}, that most of the galaxies that host
supernovae have already been photometrically observed with homogeneous
presetted criteria, so that one can study the properties of those
galaxies and correlations between the supernovae and their host
galaxies. The disadvantage is that the SDSS is somewhat too shallow
for this purpose due to a limitation arising from the
time-delay-and-integrate mode imaging with the 2.5 m aperture
telescope \citep{Gunn98,Gunn06}.

The SN candidates are spectroscopically followed-up with other
telescopes as much as the time allows to determine their types and
redshifts \citep{Zheng08}. After the SN light faded away spectroscopic
observations are carried out for its host galaxy to detremine redshift
of the SN Ia candidate.
In this paper we use the data from the
first year for the completeness is high
for redshift of host galaxies for the first year SNe.  We limit the
sample effectively to $z\lesssim 0.3$ for the sample
completeness. Five colour photometry had been made for host
galaxies. A $0.1L^*$ galaxy at $z \approx 0.3$ would give $g\sim
22.7$, $r\sim 21.9$, $i\sim 21.4$, $z\sim 21.1$, so that we can
sustain reasonable accuracy in SDSS photometry for the four passbands
\citep{Hogg01,Ivezic04}.  For the $u$ passband, however, a $0.1L^*$
galaxy at $z \approx 0.3$ gives $u\sim 24.3$, which is significantly beyond our
survey limit $u=22.0$, a calibre of our survey with a 2.5-m telescope,
and hence, the $u$ passband information cannot efficiently be
exploited.

In this paper, we take $H_0 = 70$ km s$^{-1}$ Mpc$^{-1}$, $\Omega_m=0.3$ and 
$\Omega_\Lambda=0.7$ for the cosmological parameters.

\section{Sample}

In the 2005 SDSS-II observation, 706 SNe Ia candidates were identified
including objects with observations at only a few epochs and 130 SNe
Ia among them are spectroscopically confirmed with additional 16
identified as spectroscopically probable SNe Ia. The latter are the
objects whose spectrum do not show significant Si II absorption-line
feature while the overall spectra match with the Ia templet. Follow-up
observations were not complete due to the time limitation while SNe
candidates would be observable. We are left with the candidates whose
light curves indicate SNe Ia but were not spectroscopically confirmed. 
Our experience with the spectroscopically confirmed sample tells us
that spectroscopic failure rate to confirm SNe Ia is only 10\% of the
light-curve selected candidates \citep{Sako08}. This significant
failure rate, however, applies to loosely selected candidates raised
for spectroscopic follow-up that use light curves only in early days
from the first detection, and it is substantially reduced for tightly
light-curve selected SN Ia candidates, such as those used in this
paper. A contamination of a few misidentified SNe Ia out of a hundred,
if any, would not spoil global statistical analyses, such as those
presented in this paper, beyond our statistical errors. Host galaxies
of those candidates were spectroscopically observed later and their
redshifts were measured. We refer to these cases as `host-$z$ measured
SNe Ia'. There are 168 objects in this category, which make our total
sample to comprise 314 SNe Ia with known redshift.

\begin{figure}[ht]
\plotone{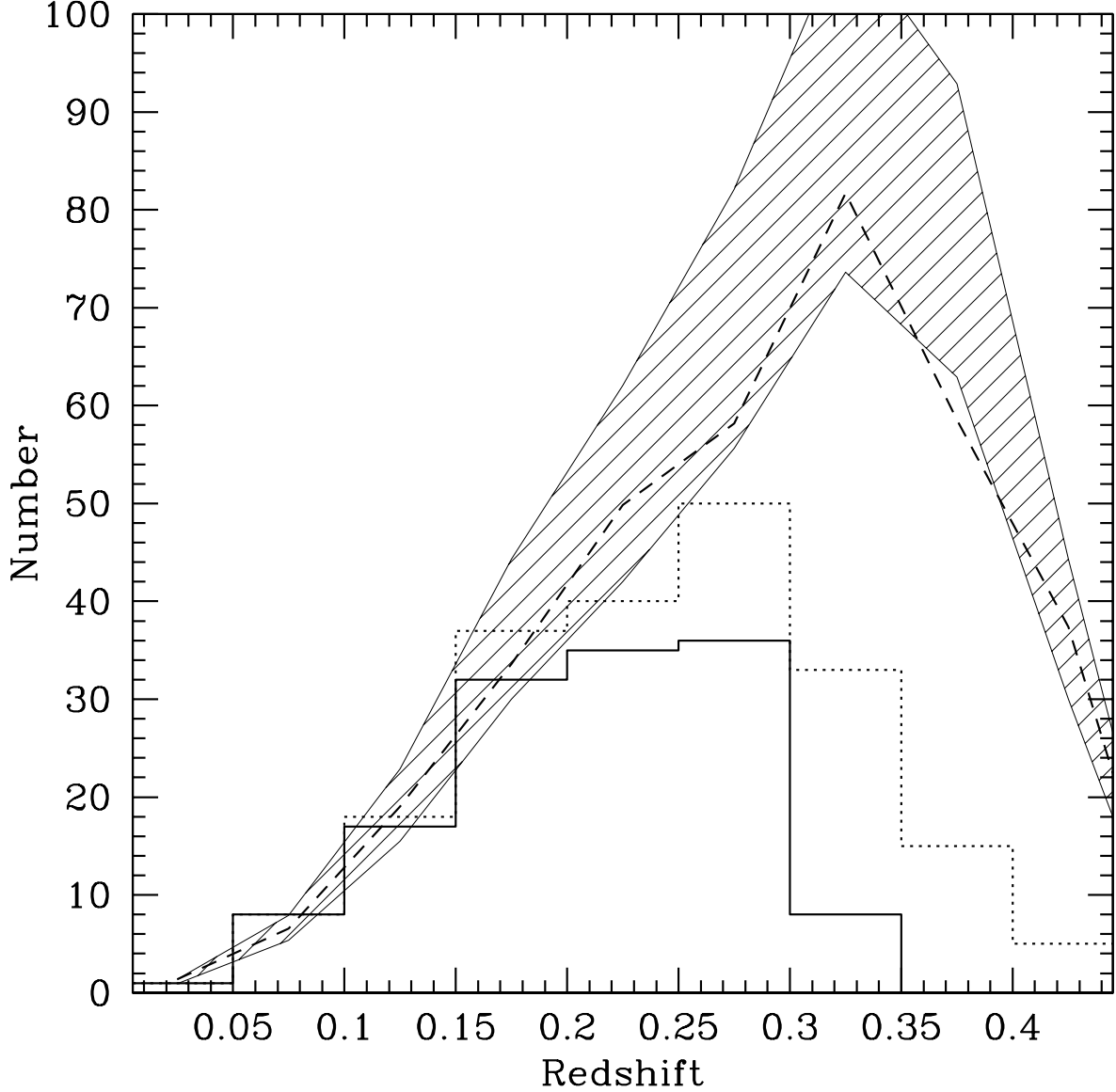}
\caption{Redshift distribution of our samples. The dotted histogram
  represents the basic sample comprising 207 SNe Ia. The solid histogram
  is the magnitude limited sample, our final sample, comprising 137 SNe
  Ia with $r_{\rm max} < 21.5$ mag used to derive the LF. Shaded region
  shows the expected distribution when SNe Ia are distributed as
  $\propto(1+z)^{1.5}$ with the selection criteria we
  adopted taken into account. Shading stands for 20\% uncertainty in the
  normalization. Dashed curve is the expected distribution when SNe Ia
  rate shows no evolution in $z$.
\label{fig:zhist}}
\end{figure}

For the supernova sample, $g$, $r$, and $i$-band light curves are
measured by fitting images with a model of the galaxy background
and the SN point spread function 
without pixel re-sampling frame by frame (scene-modeling
photometry) \citep{Holtzman08}. These SN multi-band photometric data are fit
simultaneously by the SALT2 light curve fitter \citep{SALT2}
to derive photometric parameters. 
Some SNe Ia in our sample are not suitable for
detailed analysis: for SNe Ia discovered in the early or late phase of
the observing season, i.e., early September and late November, light
curves are observed only partially, with which it is often difficult
to determine the accurate light curve shape and hence peak brightness. 
Some SNe Ia are faint and their signal-to-noise ratio at peak
brightness is too low due to their large distances. We set the
following criteria, as were done by \citet{Dilday08}, for the secure
acquisition of light curves: (i) at least one measurement with $T_{\rm
rest} < -2$ days; (ii) at least one measurement with $T_{\rm rest} >
+10$ days; (iii) at least five measurements with $-20 < T_{\rm rest} <
+60$ days; (iv) at least one measurement with signal-to-noise ratio
above 5 for each of $g$, $r$, and $i$; (v) reasonable light curve fit
whose reduced $\chi^2$ calculated by SALT2 is less than 3 for probable
and host-$z$ SNe.  For spectroscopically confirmed SNe Ia, the $\chi^2$
criteria was not applied.  Here $T_{\rm rest}$ is the rest-frame time
from the epoch of peak brightness in the rest-frame $B$-passband. 
222 SNe Ia among 314 passed these light curve criteria.
The known peculiar SNe Ia, SN2005gj \citep{Aldering06, Prieto07} and 
SN2005hk \citep{Phillips07} are rejected here.
We limit the sky area
to $-50.0 < \alpha < +55.0$ and $-1.25 < \delta < +1.25$ and the time
coverage to $53626$ (13 September 2005) $< t_{\rm max}({\rm MJD}) <
53691$ (11 November 2005) to avoid the edge effects. This leaves 207
SNe Ia with us, the
distribution of which is shown in Figure \ref{fig:zhist} as dotted histogram. 
We take this as the basic SN Ia sample. These
selections, up to item (v), are readily implemented in the analysis
procedure. The effective area of the survey is 262.5 square degree.

\begin{figure}[ht]
\plotone{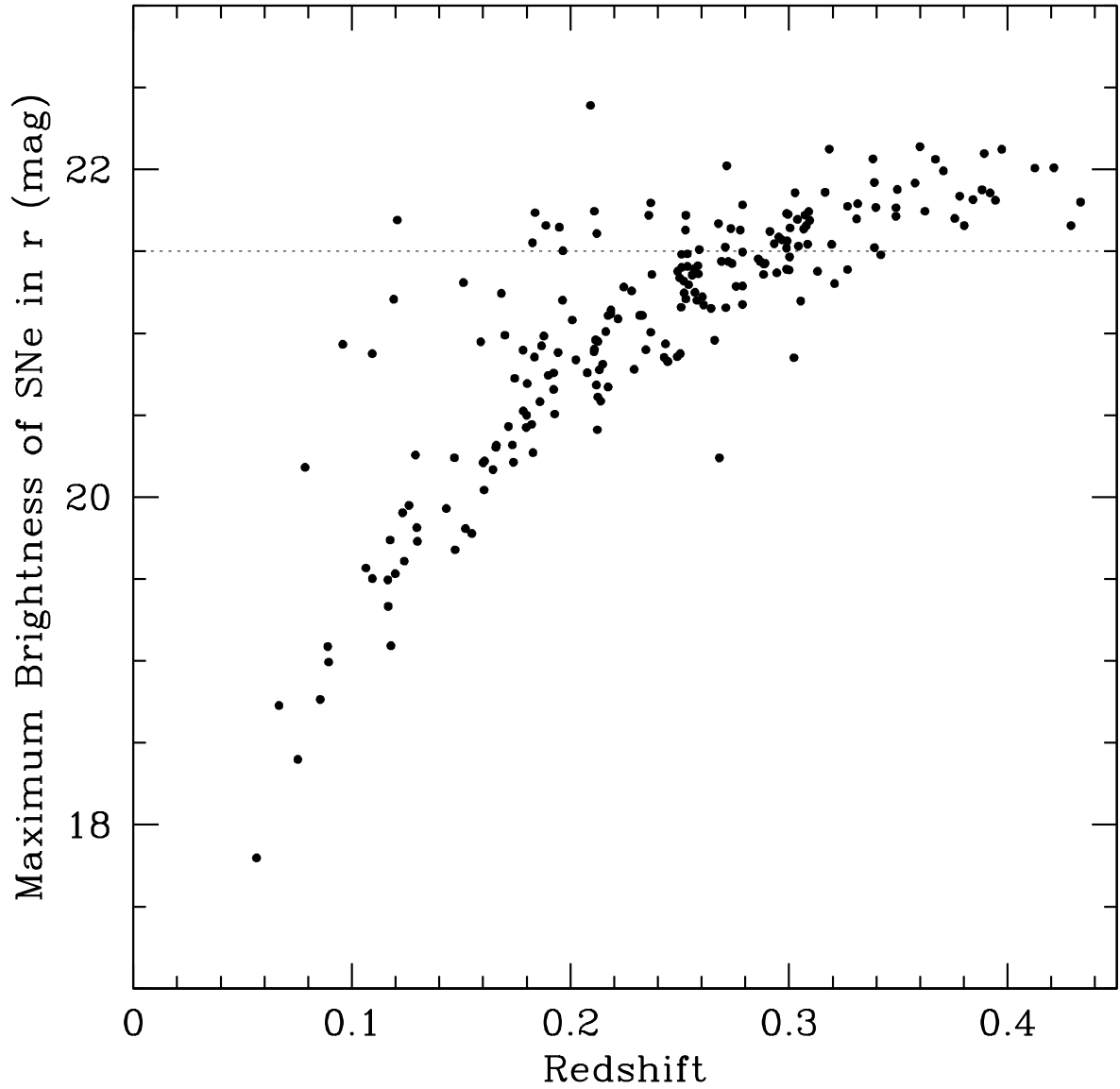}
\caption{Apparent maximum brightness of 207
SNe Ia in the $r$-band as a function of
redshift. Horizontal dotted line denotes the
magnitude limit of $r_{lim} = 21.5$ mag. Brightness
is not corrected for extinction or colour excess.
\label{fig:zrmax}}
\end{figure}

To make the sample incompleteness well defined for distant faint
supernovae and the sample suitable to an application of the $1/V_{\rm
max}$ method to compute the LF, we make the sample magnitude limited.  
Figure \ref{fig:zrmax} shows apparent maximum
brightness in the $r$-passband as a function of redshift, which suggests
us to set the limiting magnitude to be $r_{\rm lim} = 21.5$mag. The
figure shows that this ensures reasonable completeness of SNe Ia at $z
= 0.20 - 0.25$.  After applying this limiting magnitude, 137 SNe Ia
are left with us for a magnitude limited sample, with the mean
redshift $\langle z\rangle=0.207$, which we take as our final sample
for our analysis. Among magnitude limited sample, 72 are
spectroscopically confirmed, 5 are spectroscopically probable and 60
are host-$z$ SNe.  Summary of the sample selection is shown in Table
\ref{tbl:sample}.

The solid histogram in Figure \ref{fig:zhist} represents the redshift
distribution of this magnitude limited sample.  The shade indicates
the distribution expected for the empirical redshift distribution 
calculated using the code to generate llight curves 
SNANA\footnote{\tt http://sdssdp47.fnal.gov/sdsssn/SNANA-PUBLIC/} 
(R. Kessler et al. 2009, private communication, submitted to PASP) assuming
the SN Ia rate, $r_{\rm SNe Ia}\propto (1+z)^{1.5\pm0.6}$, which is inferred
from the observed SDSS SNe Ia rate at low redshift \citep{Dilday08}.
The conditions of the SDSS observations, the software search
efficiency and the light
curve criteria are taken into account, and the normalisation is determined from
the number of SNe below $z=0.15$, for which redshift range SDSS SN
observation is complete.  The detail of simulation is described in
Appendix.  The shade stands for $\pm20$\% uncertainties of
the normalization, corresponding to the Poisson error of the number of SNe
Ia below $z=0.15$. The SN frequency expected for 
no evolution of SN rate is shown by the dashed curve, which lies within
the error indicated by shades.
This figure shows that
our sample is likely complete to $z < 0.20$.

\begin{figure}[ht]
\plotone{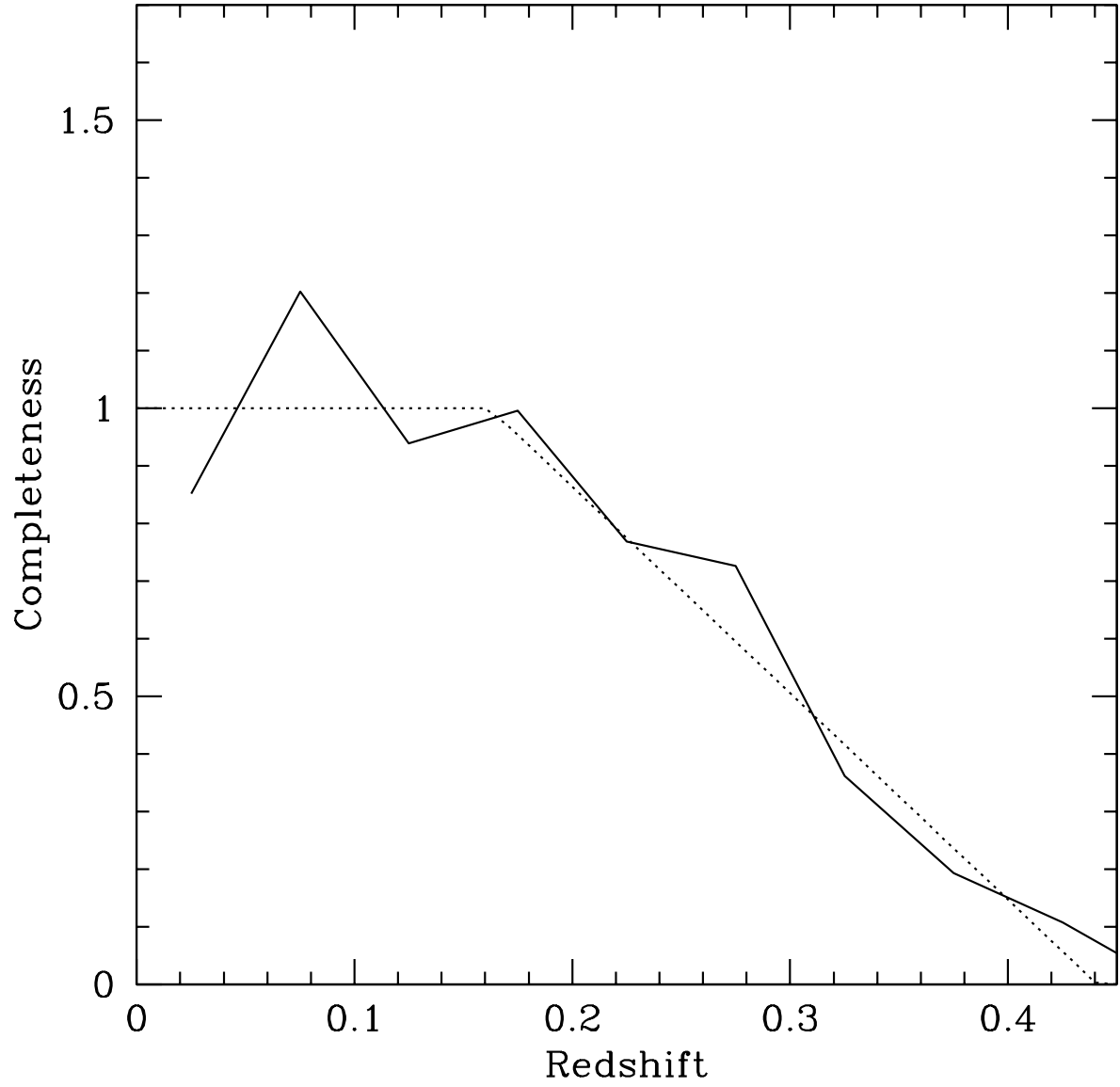}
\caption{Solid line shows the ratio of the number of SNe Ia in our
basic sample 
to the simulation as a function of redshift. Dotted line shows 
the completeness function eq.(\ref{eqn:completeness}) used in this paper. 
\label{fig:completeness}}
\end{figure}

The sample incompleteness at higher redshift is caused mainly by 
spectroscopic
targetting and is given by fitting
the ratio of observed number of SNe in our
basic SN Ia sample to the simulated number as seen in Figure
\ref{fig:completeness}:
\begin{equation}
\epsilon(z) = \left\{
\begin{array}{ll}
1.0 & z \leq z_c \\
1.0 - (z - z_c) / \Delta z & z_c < z < z_c  + \Delta z \\
0.0 & z \geq z_c + \Delta z, 
\end{array}
\right.
\label{eqn:completeness}
\end{equation}
where the best fit values are $z_c = 0.162 \pm 0.029$ and 
$\Delta z = 0.279 \pm 0.041$. 

\begin{figure}[ht]
\plotone{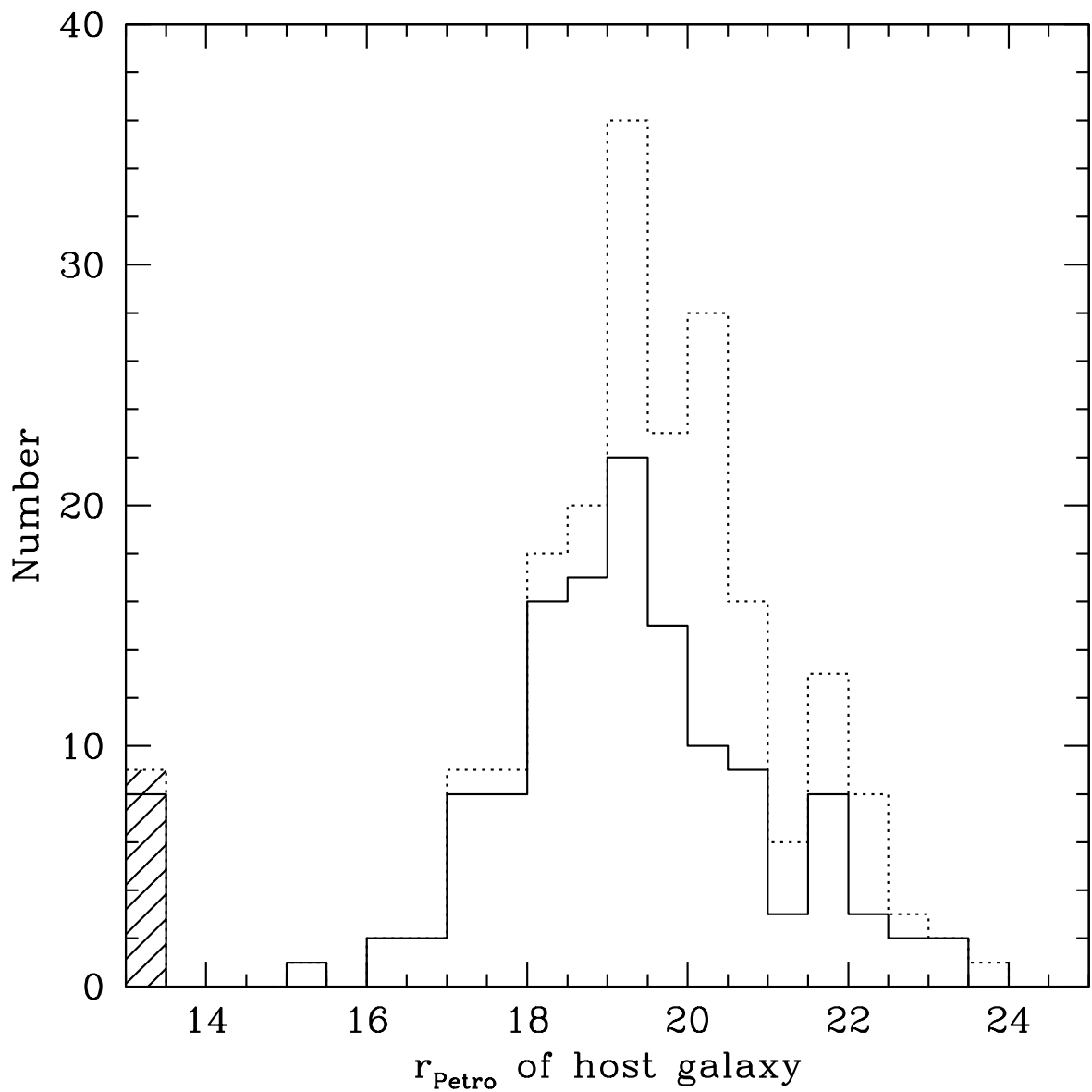}
\caption{Distributions of apparent $r$-band Petrosian magnitude of host
galaxies of our basic SN Ia sample (dotted histogram) and of the 
magnitude limited sample (solid histogram).
There are 9 (or 8) SNe Ia whose host galaxies are not identified in
the two samples, 
which are indicated in the leftmost bin with shades.
\label{fig:hrhist}}
\end{figure}

For each SN Ia the nearest primary object that resides within 10
arcsec from the SN is identified as its host galaxy in Catalog Archive
Server of SDSS Data Release 6 (\citealt{DR6}; see also
\citealt{Stoughton02}), with the identification visually
confirmed. When misidentification is suspected, it arises mostly from
deblended galaxies. Figure \ref{fig:hrhist} shows the apparent
$r$-band Petrosian magnitude distribution of host galaxies for 207 SNe
Ia before the magnitude cutoff with the dotted histogram.
The solid histogram is the host
galaxies for our magnitude limit sample. There are 9 SNe whose host
galaxies are not identified in the basic sample of 207 SNe (8 SNe in
the 137 magnitude limit sample) within $\sim10$ arcsec from a
SN. These hostless SNe Ia are listed in Table \ref{tbl:hostless},
which gives the upper limit on absolute brightness of host
galaxies searched in our survey to be $0.025-0.09L^*$ for the apparent
magnitude limit of $r=22.2$ mag. The physical distance
corresponding to 10\arcsec~ is also given in the table. Petrosian
magnitudes of $u$, $g$, $r$, $i$, and $z$-passband are used to compute
rest-frame absolute brightness of host galaxies in 5 passbands using
{\tt kcorrect v4\_1\_4} \citep{kcorrect}. Redshifts are fixed to the
spectroscopic values.

\section{Application of the $1/V_{\rm max}$ method}
\label{sec:analysis}

The light curves of SNe are fit by the SALT2 \citep{SALT2}, which
yields apparent maximum brightness in the rest-frame $B$-passband, $m_B$,
together with the parameter describing the spectral variation $x_1$
and the optical depth
representing the colour excess at maximum brightness $c$. The
parameter $x_1$ is the weight of the next leading component of the
spectral templet and is related to the light curve shape parameter
$\Delta m_{15}$ (Phillips 1993), 
as $\Delta m_{15}=1.09-0.161x_1+0.013x_1^2+O(x_1^3)$
\citep{SALT2}. SALT2 corrects for Galactic extinction following the
extinction map of \citet{Schlegel98}.

We apply the $1/V_{\rm max}$ method to calculate LFs of SNe Ia (and of
their host galaxies) to the magnitude limit sample of 137 SNe whose
maximum brightness is brighter than 21.5 mag in the $r$-passband in the
observed frame, by calculating the maximum redshift, $z_{\rm max}$, at
which each SNe Ia (or their host galaxies) 
can be observed within the magnitude limit. 
Since the multicolour passband light curve fit
by SALT2 gives us the spectral energy distribution of SN Ia,
characterised with the shape parameter and the colour excess
parameter, we can calculate expected apparent brightness in the $r$-passband
in the observed frame at various redshifts using this spectral energy
distribution.  The limiting redshift gives the maximum volume that can
be surveyed, $V_{\rm max}$, as

\begin{equation}
V_{\rm max}^i = \frac{\omega}{4\pi}\int_0^{z_{\rm max}^i}\frac{dV}{dz}(z)dz,
\end{equation}

\noindent
where $\omega$ is the solid angle 
$105\times2.5=262.5\sq\arcdeg$, and the index $i$ refers to SNe Ia.
Since $V^i_{\rm max}$ is regarded as the volume surveyed to find
$i$-th SNe, the LF is obtained by summing the
inverse of $V_{\rm max}^i$ within specified magnitude bins ($\Delta M$)
assuming that the LF is not evolving over the respective redshift range.
Taking into account the effect of visibility time and spectroscopic
incompleteness, the LF is calculated as

\begin{equation}
\phi(M)\Delta M ({\rm Mpc}^{-3}) = \sum_{i \in |M^i-M| < \Delta M/2} \frac{\tau}{V_{\rm max}^i \times ct^i \times \epsilon(z^i)},
\label{eqn:analysis}
\end{equation}

\noindent
where $ct^i$ is the time of visibility in the rest-frame at which each
SN Ia is observed. If the SN would be observed just at one epoch, as
was done in a number of observations to derive the SN rate, the
visibility time will be a time span over which each SNe can be
detected above the detection limit: in this case a fainter SNe would
have a shorter visibility time.  In our case, however, observations
have been made for the same field of sky continuously with the
magnitude limit set for peak brightness.  The visibility time will
then be a time span of the survey observation.  From the criteria on
the date of maximum brightness, the visibility time in the observed
frame is 65 days, and hence $ct^i = 65 / (1+z^i)$ days in the
rest-frame. For SNe Ia $M^i$ is absolute peak $B$-passband magnitude of SNe Ia,
whose apparent rest-frame magnitude is estimated from SALT2.  For
host galaxies it is absolute magnitude in the rest-frame estimated using {\tt
kcorrect}.  The factor $\epsilon(z)$ is the completeness correction,
eq. (\ref{eqn:completeness}). The LF is represented in units
of per Mpc$^3$, and $\tau$ (year) absorbs the time of the duration of
observability when we deal with SNe. We take $\tau = 1$ yr as the
unit.
We describe in Appendix simulations we
made to show that sample incompleteness and our corrections do not
induce particular systematic errors to our analysis and our procedures
allow us to recover the true LF, SN rate and related quantities.

\section{Luminosity function of SNe Ia}

\begin{figure}[ht]
\plotone{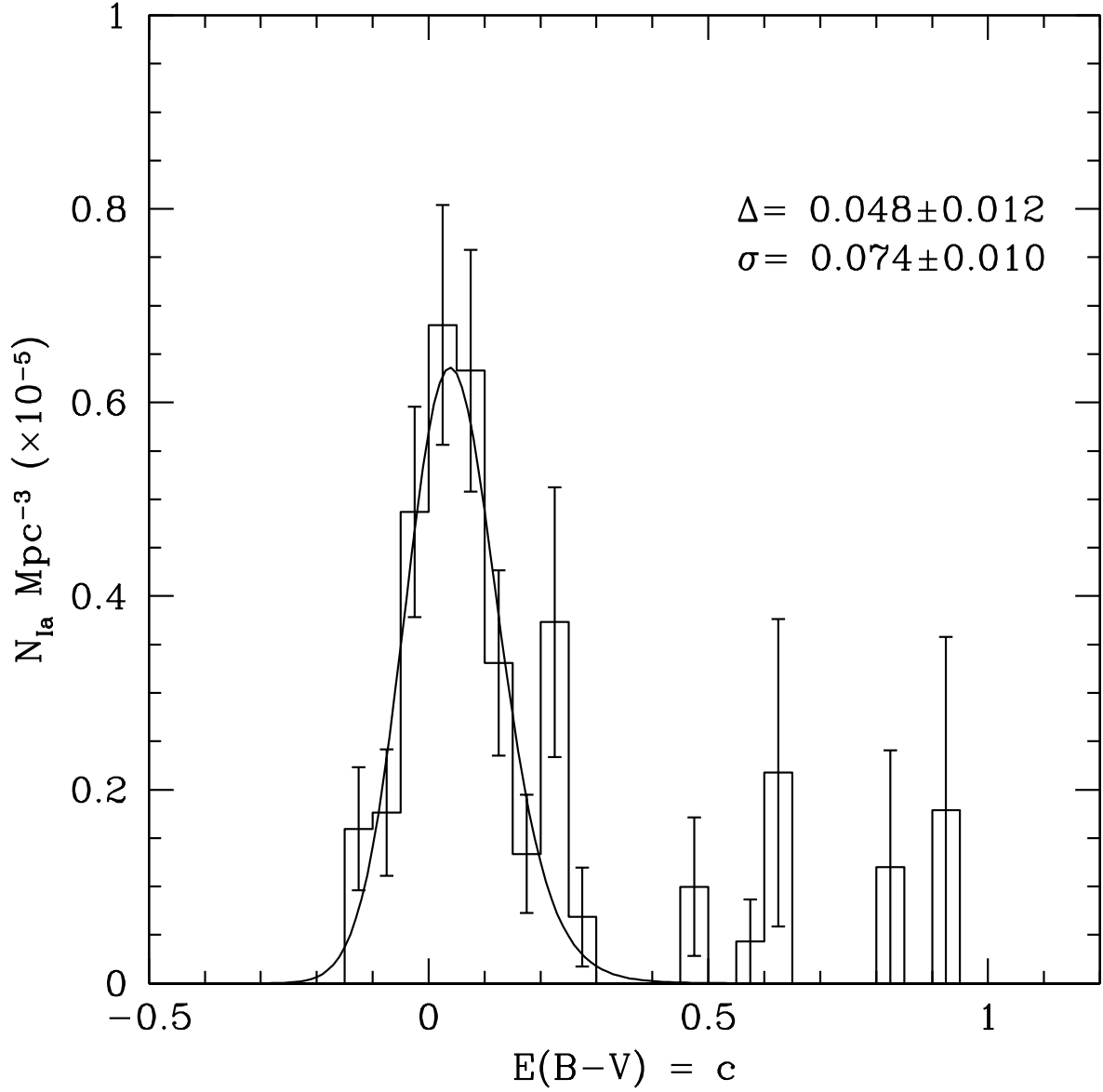}
\caption{Distribution of colour excess $E(B-V) = c$
calculated from the magnitude limited sample.
The curve is the Gaussian with the function $p(E(B-V))
\propto \exp (-E(B-V)/\Delta)$ for $E(B-V)>0$ and $p(E(B-V))=0.0$ for
$E(B-V)<0$ convolved with Gaussian (see text).
\label{fig:ebvhist}}
\end{figure}

To estimate the intrinsic brightness of SNe Ia, we must correct for
dust extinction within host galaxies.
In fact, SNe Ia show the variation in colour that could be attributed to dust
extinction within host galaxies and/or interpreted as an intrinsic
colour variation. The colour information is obtained from the colour
excess parameter $c$ of SALT2, which is defined by $c = (B-V)_{\rm
max} - \langle(B-V)_{\rm max}\rangle$ at $B$-passband maximum
brightness, where the second term is colour of the SN Ia templet.
Figure \ref{fig:ebvhist} shows the distribution of this colour excess
$c=E(B-V)$.  The distribution is asymmetric with respect to $E(B-V) =
0$, similar to that of the SNLS sample \citep{SNLS}. If we assume the
color distribution as an exponential distribution $\propto
\exp(-E(B-V)/\Delta)$ smeared by intrinsic Gaussian color distribution
with the dispersion of $\sigma$, the observed distribution can be
fitted with $\Delta = 0.048$ and $\sigma = 0.074$ as in Figure
\ref{fig:ebvhist}. This value of $\Delta$ is smaller than $\Delta =
0.138$ obtained for the nearby SN Ia sample by \citet{Jha07}.  The
mean and dispersion $\langle c\rangle = 0.176\pm 0.280$ are compared
with $\langle E(B-V)\rangle = 0.128 \pm 0.173$ from \citet{Jha07}.
Several data points with high value of $E(B-V)$ greater than 0.5
contribute significantly to $c$.  If they would be removed the mean
and dispersion will become $\langle c\rangle = 0.061\pm 0.107$.  One
may suspect that our sample may lack highly extinct or very red SNe
due to selection effects. It is unlikely, however, that we dropped
such SNe at least at lower $z$ where SNe are surveyed deep enough to
reach brightness significantly fainter than are dimmed by extinction,
unless $E(B-V)>1$ which are dropped as our selection procedure for the
SN Ia sample.  The fact that the distribution does not change
appreciably from low $z$ to our redshift limit indicates that the
decline of the event rate that give $E(B-V)>0.2$ is not due to the
selection effect and SNe that give a larger $E(B-V)$ is missing.  The
distribution of $E(B-V)$ corresponds, if interpreted as extinction, to
extinction in the $V$ passband ranging from $-0.14$ mag to $0.53$ mag
for one sigma, if $R_V=3.1$ is assumed. The range of positive numbers
is consistent with what is expected from extinction in the galactic
discs with the HI column density $N_{\rm HI}\lesssim 9\times 10^{20}$
cm$^{-2}$ \citep{BH78}, which is a typical value for disc galaxies
unless they are edge-on. Extinction may also receive a contribution
from SN itself and its environment. We come back to this problem in
the end of the next section.  

\begin{figure}[ht]
\plotone{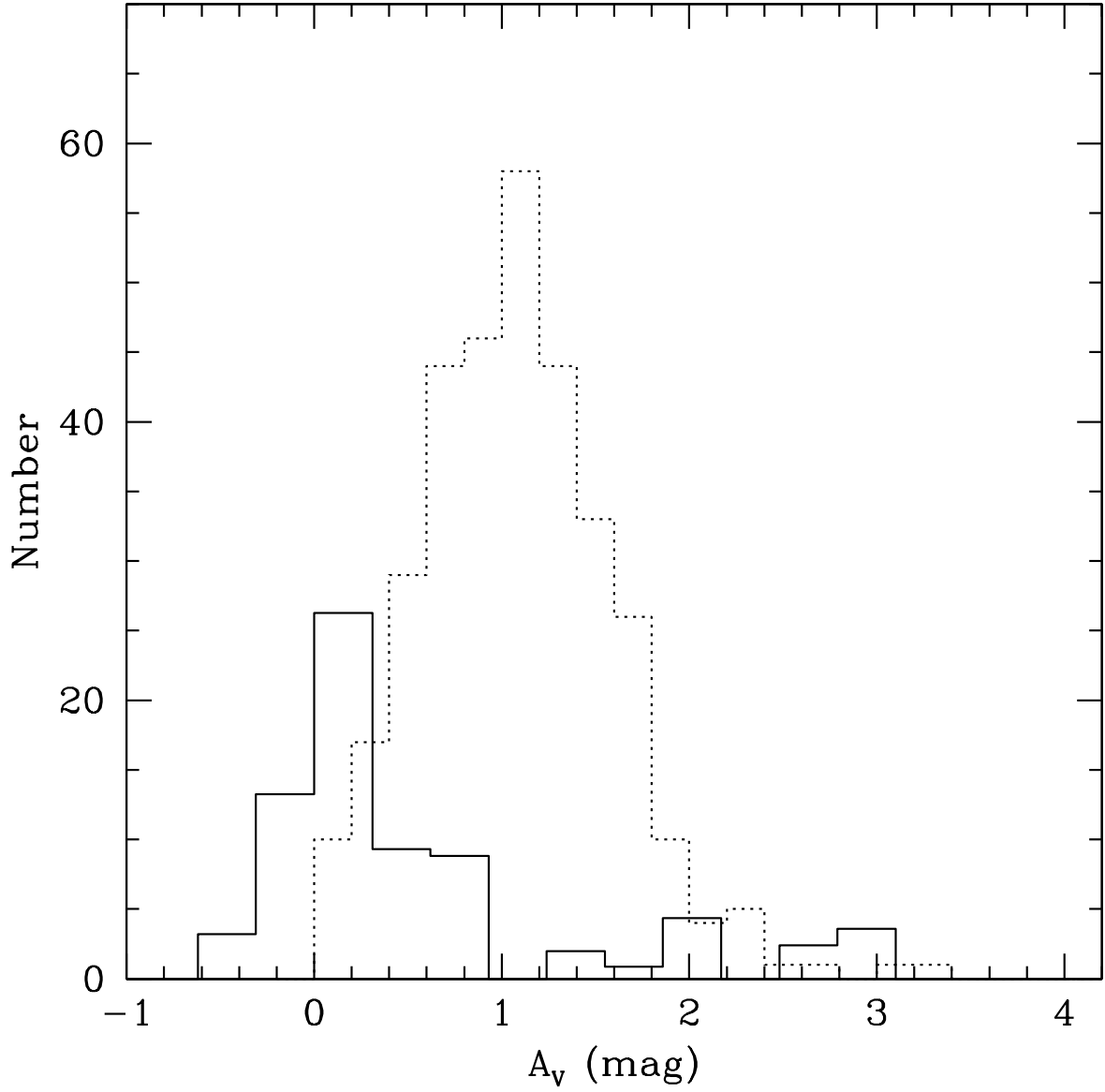}
\caption{Distribution of $V$-band extinction $A_V$ for SNe Ia, 
estimated assuming that 
their colour
excess is due to dust reddening and $R_V=3.1$
(solid histogram).
The dotted histogram is $A_V$ of field galaxies 
inferred from the Balmer emission line ratio H$\alpha$/H$\beta$ 
\citep{Nakamura04} for comparison.
\label{fig:Av}}
\end{figure}

Figure \ref{fig:Av} shows the distributions of $A_V$ for the SNe Ia
sample, taking the colour variation as reddening and assuming $R_V=3.1$.
Note that above $A_V=3$ SNe Ia candidates, if any, are rejected during SN
identification with light curves \citep{Sako08}.  This is compared
with $A_V$ for star forming field galaxies (dotted histogram), as
calculated from the H$\alpha$/H$\beta$ Balmer line ratio obtained by
SDSS within 3 arcsec of the central region of galaxies for $z<0.2$
\citep{Nakamura04}.  The mean value of extinction of SNe Ia, 
$\langle A_V\rangle=0.20$ mag,
is significantly smaller than that for star forming field galaxies,
$\langle A_V\rangle=1.10$ mag.  This suggests that most SNe Ia take
place in regions clearer in dust than actively star forming regions.

We provisionally assume that the colour variation is due to
extinction within host galaxies and that the extinction law is the
same as that in our Galaxy with $R_V = 3.1$. We then estimate
brightness of SNe Ia by correcting for extinction as $M_B - \beta
\times c$ with $\beta =R_V+1=4.1$. We note, however, that this ratio
of total to selective extinction is not well justified. There are some
indications that $R_V$ for SNe Ia is lower than the Galactic value;
\citet{Altavilla04} used the bluest SNe Ia to
estimate intrinsic colour of SN Ia and \citet{Reindl05} used SNe
Ia in early-type galaxies and far outlying SNe Ia in spiral
galaxies. Both authors obtained $R_V \sim 2.5$. 
\citet{Nobili08} also found a low value of $R_V = 1.75$
simultaneously deriving templets of SNe colour evolution in time 
and extinction.
\citet{SNLS} resulted in $R_V = 0.57$ and 
\citet{Kessler09} gave $R_V=2.18$. 
All $R_V$ thus derived are smaller than the canonical value
for the Milky Way.

Figure \ref{fig:snlf} shows the LF of SNe Ia in the $B$-passband, which is 
fitted with the Gaussian distribution with the mean and
dispersion given in the upper left corner of each panel.
The panel (a) assumes that the colour variation $c$ arises from
extinction with $\beta =4.1$, giving 
the mean $M_B^0 = -19.42$ (in the Johnson zero point) and 
the dispersion $0.24$ mag.
If we would adopt a smaller value for $R_V$, the LF
of SNe Ia becomes closer to a more regular Gaussian distribution with
a smaller dispersion, as shown in Figure \ref{fig:snlf}(b).  The
minimisation of dispersion with respect to $\beta$ results in
$\beta=2.93$, which leads to the narrower Gaussian width of 0.16 mag
with $M_B^0= -19.32$, which is 0.10 mag fainter than with $\beta=4.1$.
Thus the smaller
$R_V$ gives more homogeneous SN Ia
luminosity.

This small value of $R_V$ obtained by minimizing the width of Gaussian
fitted to SNe LF is an alternative manifestation of 
the low value of $\beta$
obtained by \citet{SNLS} by
minimising the residual scatter in the Hubble diagram along with
cosmological parameters.
\citet{Kowalski08} and \citet{Kessler09} indicated that minimizing
the scatter in the Hubble diagram tends to give $R_V$ biased towards a value 
lower than the
true value. Our simulation, however, shows that the value of $\beta$ 
may be biased to the lower value but no more than
by $\sim0.1$, so that this cannot be the reason for 
$R_V$ being significantly smaller than 3.1.
We do not conclude here that $R_V$ is actually smaller but
take $\beta=4.1$ as our fiducial choice, keeping in mind the uncertainty 
from $R_V$ in the analysis in what follows.

It has been argued that the maximum luminosity correlates with the
light curve shape, or more specifically the decline rate ($\Delta
m_{15}$, stretch, or SALT2's $x_1$), and the inclusion of the
correlation with it makes the behaviour of the LF
tighter \citep{Pskovskii84,Phillips93,Hamuy96}.  We show in panels (c)
and (d) of Figure \ref{fig:snlf} the LF where brightness is corrected
by $\alpha x_1$, with $\alpha$ chosen to minimise the width of the
Gaussian. In panel (c), $\beta$ is fixed to our fiducial value of 4.1
and the minimisation gives  $\alpha = 0.052$. In (d)
both $\alpha$ and $\beta$ are chosen
to minimise the width, which results in $\beta = 2.52$ and $\alpha =
0.123$.  
The correlation of $x_1$ with
brightness makes the Gaussian distribution narrower, especially for
the case in which both $\alpha$ and $\beta$ are optimised.
The narrowest Gaussian is obtained with $\alpha=0.123$ which corresponds
to $\alpha^\prime=0.76$ where the correction for the light curve shape to
brightness is
expressed as $\alpha^\prime \Delta m_{15}$. This value is consistent
with 0.78$\pm0.18$ obtained by \cite{Hamuy96}.

\begin{figure}[ht]
\plotone{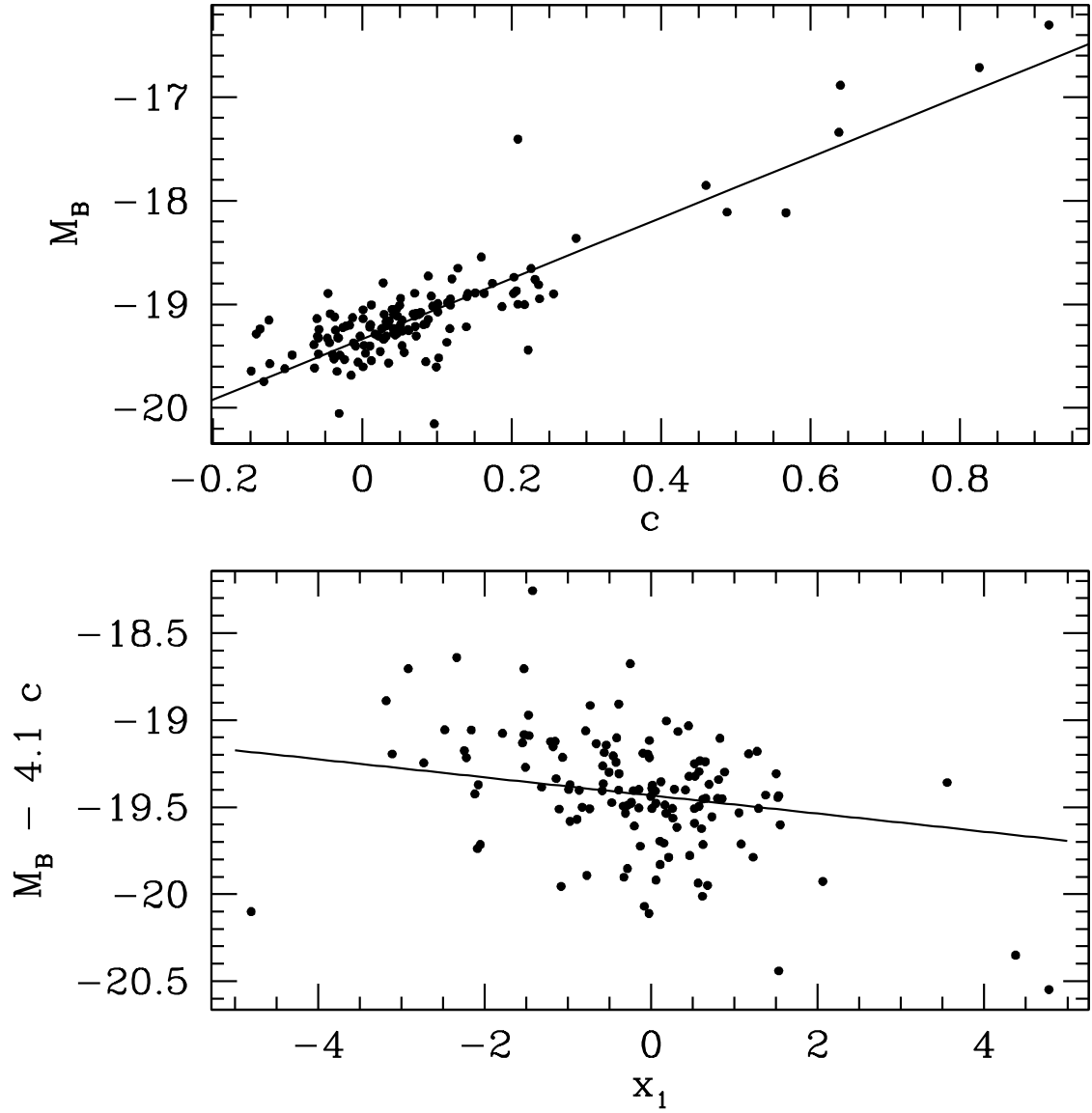}
\caption{Correlations between brightness of SNe and the colour excess 
parameter $c$ (the top panel) and the $x_1$ parameter (the bottom panel). 
\label{fig:Mab}}
\end{figure}

Figure \ref{fig:Mab} shows the correlation between brightness of SNe
with the shape parameter and the colour excess. This correlation is the reason
why the width of the luminosity function decreases upon inclusions of
the light curve shape parameter and the colour excess parameter 
(with a small $R_V$). The correlation in the upper panel is represented by
$M_B=-19.34+2.93c$ as shown in Figure \ref{fig:snlf}(b), and
that in the lower panel by
$M_B-4.1c=-19.43-0.052x_1$ as in Figure \ref{fig:snlf}(c).

\section{Host galaxies}

\begin{figure}[ht]
\plotone{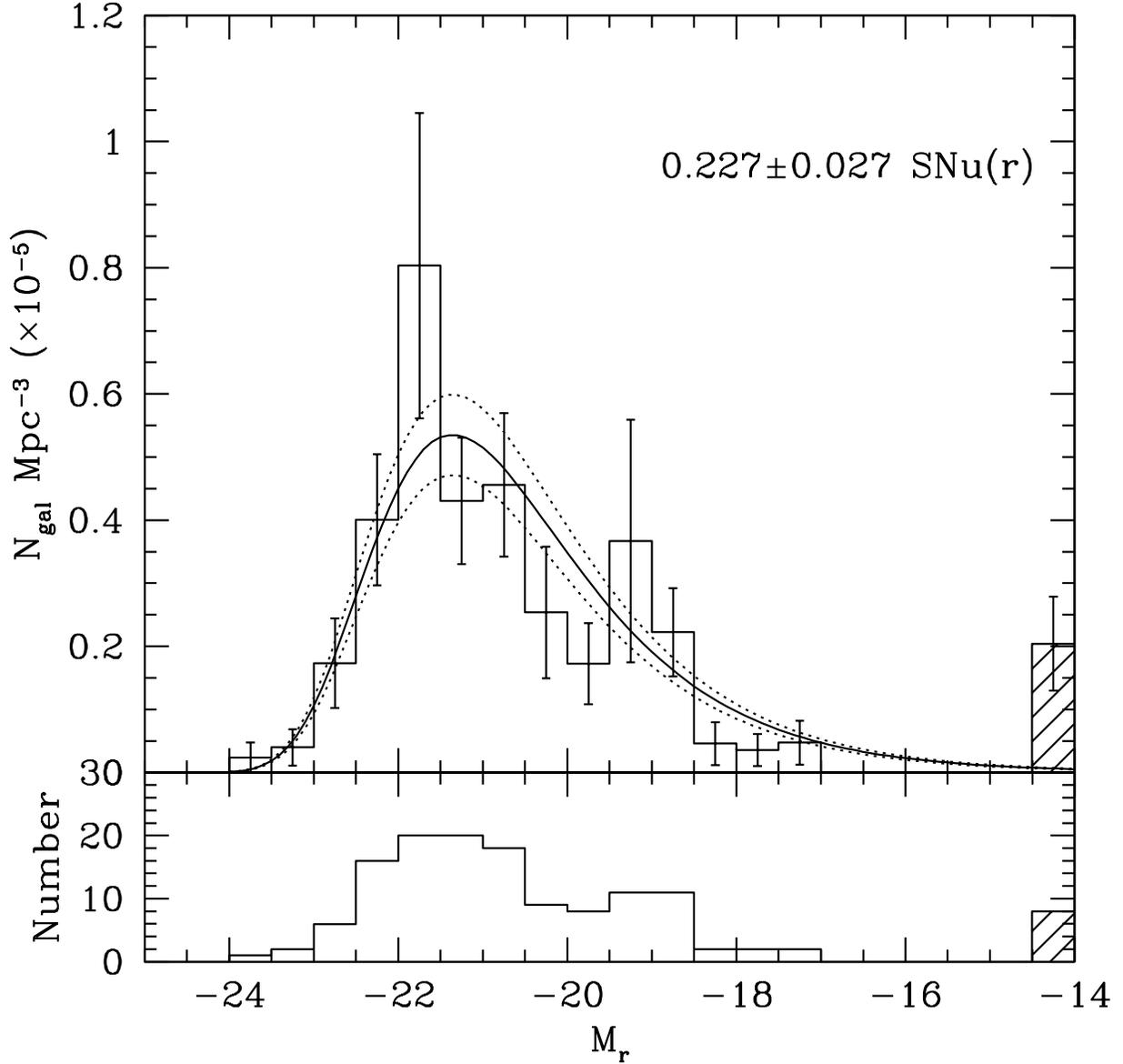}
\caption{Luminosity function of SNe Ia host galaxies in the $r$-passband. Lower
panel shows the number of contributing galaxies in each bin. The solid curve is
the luminosity weighted luminosity function of field galaxies 
\citep{Blanton01}
normalized to fit the histogram. 
The scale factor that gives the supernova
rate in SNu is given at the top right of the figure. 
Dotted lines show the range corresponding to the fitting error.
Contribution from hostless SNe are indicated in the rightmost bin with shades.
\label{fig:hostlf_r}}
\end{figure}

\begin{figure}[ht]
\plotone{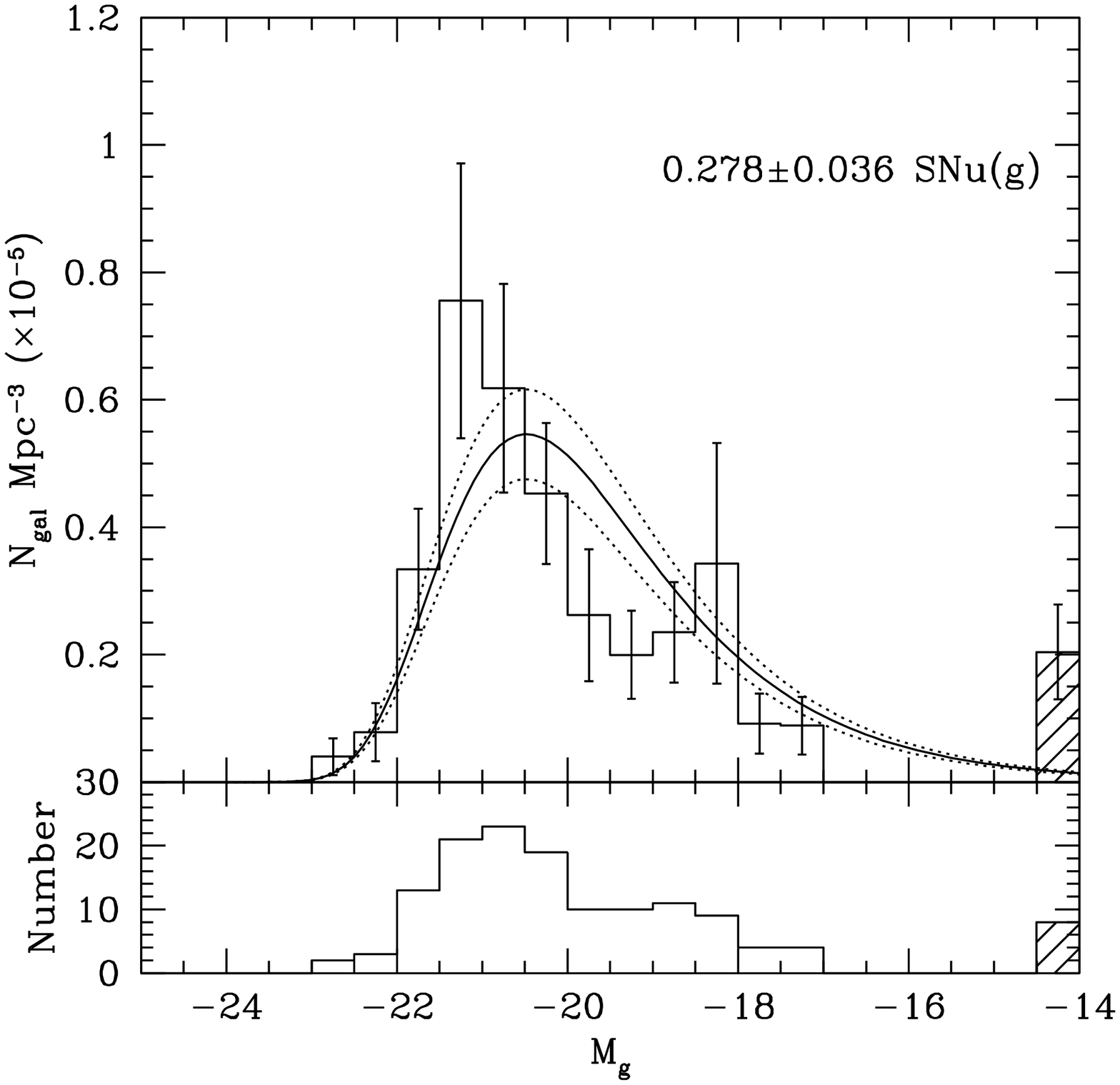}
\caption{Same as figure \ref{fig:hostlf_r} but in the $g$-passband.
\label{fig:hostlf_g}}
\end{figure}

\begin{figure}[ht]
\plotone{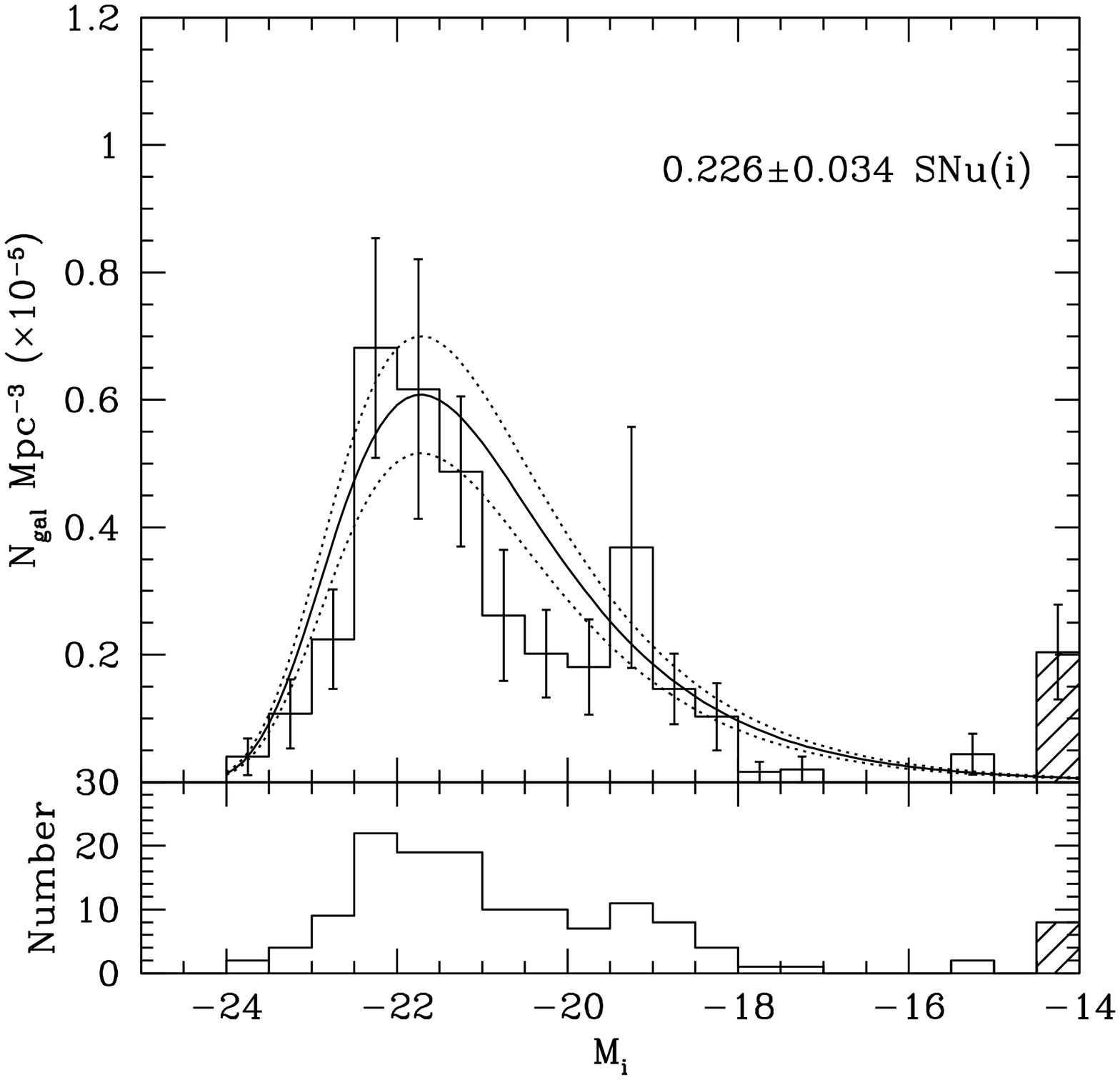}
\caption{Same as figure \ref{fig:hostlf_r} but in the $i$-passband.
\label{fig:hostlf_i}}
\end{figure}

The LFs of SNe Ia host galaxies, as calculated by the $1/V_{\rm max}$
method, are shown in Figures \ref{fig:hostlf_r}, \ref{fig:hostlf_g}
and \ref{fig:hostlf_i} with solid histograms for the $r$, $g$, and
$i$-passbands. Hostless SNe are indicated by shaded histogram at
rightmost bin.
We draw with solid curves the LF of general field galaxies
obtained by \citet{Blanton01}, multiplied with the luminosity. The
curves show good match of the LF of SN host galaxies with that of the
field galaxies, which means that the LF of galaxies derived from SNe
Ia faithfully represents that of galaxies in the field. 
We do not see
any particular deviations between the two for three colour passbands 
$g$, $r$ and $i$ we studied, meaning that the occurrence
of SNe Ia is primarily proportional to the luminosity of galaxies. 
Matching the
two LF's gives the SNe Ia rate in the
conventional supernova unit (SNu), the SN rate per $10^{10}$ solar
luminosity per century $r_L$.
Taking $M_r(\sun)=4.62$, we obtain 
\begin{equation}
r_L=0.227\pm0.027\;{\rm SNu}(r),
\end{equation}
where we use the luminosity normalisation of \citet{Blanton03}
shifted to $z=0.2$ using their evolution prescription
with the $Q$ parameter that represents the evolution of
luminosity per redshift interval\footnote{
The normalisation of the luminosity functions of
\citet{Blanton03} differs significantly from \citet{Blanton01} 
apart from the use of the different passbands defined at $z=0.1$: 
the luminosity density of the latter
is 20-40\% larger than the former.
The difference in the shape of the luminosity function, however, is modest and 
we adopt Schechter function parameters of \cite{Blanton01} for the luminosity 
function in the $g,r,i$ passbands at $z=0$ by
shifting only the normalisation.
For the $g$, $r$, and $i$ passbands, we interpolated across the five colours
at $z=0.1$.
The luminosity densities at $z=0.2$ for the $g$, $r$, and $i$ passbands adopted
are ${\cal L}_g=1.39\times10^8L_\odot/({\rm Mpc})^3$, 
${\cal L}_r=1.60\times10^8L_\odot/({\rm Mpc})^3$, 
${\cal L}_i=1.87\times10^8L_\odot/({\rm Mpc})^3$}. 
This matching was done using data points at magnitude bins where the number
of contributing galaxies are greater than 10 and error is 
Poisson from the SN number used in the $1/V_{\rm max}$ analysis. 
The rate given above can be treated as
a mean over the redshift range of our sample, the mean redshift
$z=0.20$.
We may convert this rate expressed in SNu to 
the volumetric rate by multiplying the luminosity density.
With reference to the luminosity density of \citet{Blanton03}, we find the
volumetric rate of SNe Ia,  
\begin{equation}
r_V=(3.63\pm0.43)\times10^{-5}\;{\rm Mpc}^{-3}\;{\rm yr}^{-1}. 
\label{eq:volsnr}
\end{equation}
For $g$ and $i$ passbands the matching of the two LF's yield
$r_L=0.278\pm 0.036\;{\rm SNu}(g)$ and $r_L=0.226\pm 0.034\;{\rm SNu}(i)$
taking $M_g(\sun)= 5.07$ and $M_i(\sun)= 4.52$ mag, respectively. 
These SN rates yield the volumetric rate 
$r_V=(3.86\pm0.50)\times10^{-5}\;{\rm Mpc}^{-3}\;{\rm yr}^{-1}$
for the $g$ passband and 
$r_V=(4.23\pm0.64)\times10^{-5}\;{\rm Mpc}^{-3}\;{\rm yr}^{-1}$ for
the $i$ passband. The rates obtained from the three passbands agree
with each other, within a 1 $\sigma$ error, showing the internal 
consistency.  
These SN rates agree with 
$r_V=(3.70\pm0.41\pm0.34)\times10^{-5}\;{\rm Mpc}^{-3}\;{\rm yr}^{-1}$ 
obtained directly by 
summing up the LF of SNe Ia in Figure \ref{fig:snlf}(a).
Here the second error represents that arising 
from the completeness function given in eq. (\ref{eqn:completeness}),
which depends on the SN rate normalisation.
If we take the $B$-band luminosity density
${\cal L}_B=1.44\times10^8\;L_\sun^B\;{\rm Mpc}^{-3}$ 
at $z=0.2$, obtained by interpolating across five colours,
we obtain the supernova rate per $B$-band luminosity 
\begin{equation}
r_L=0.257\pm 0.028\pm0.024\;{\rm SNu}(B).
\end{equation}
This is compared with earlier measurements at low redshift 
compiled in
Table \ref{tbl:snrate}.

\begin{figure}[ht]
\plotone{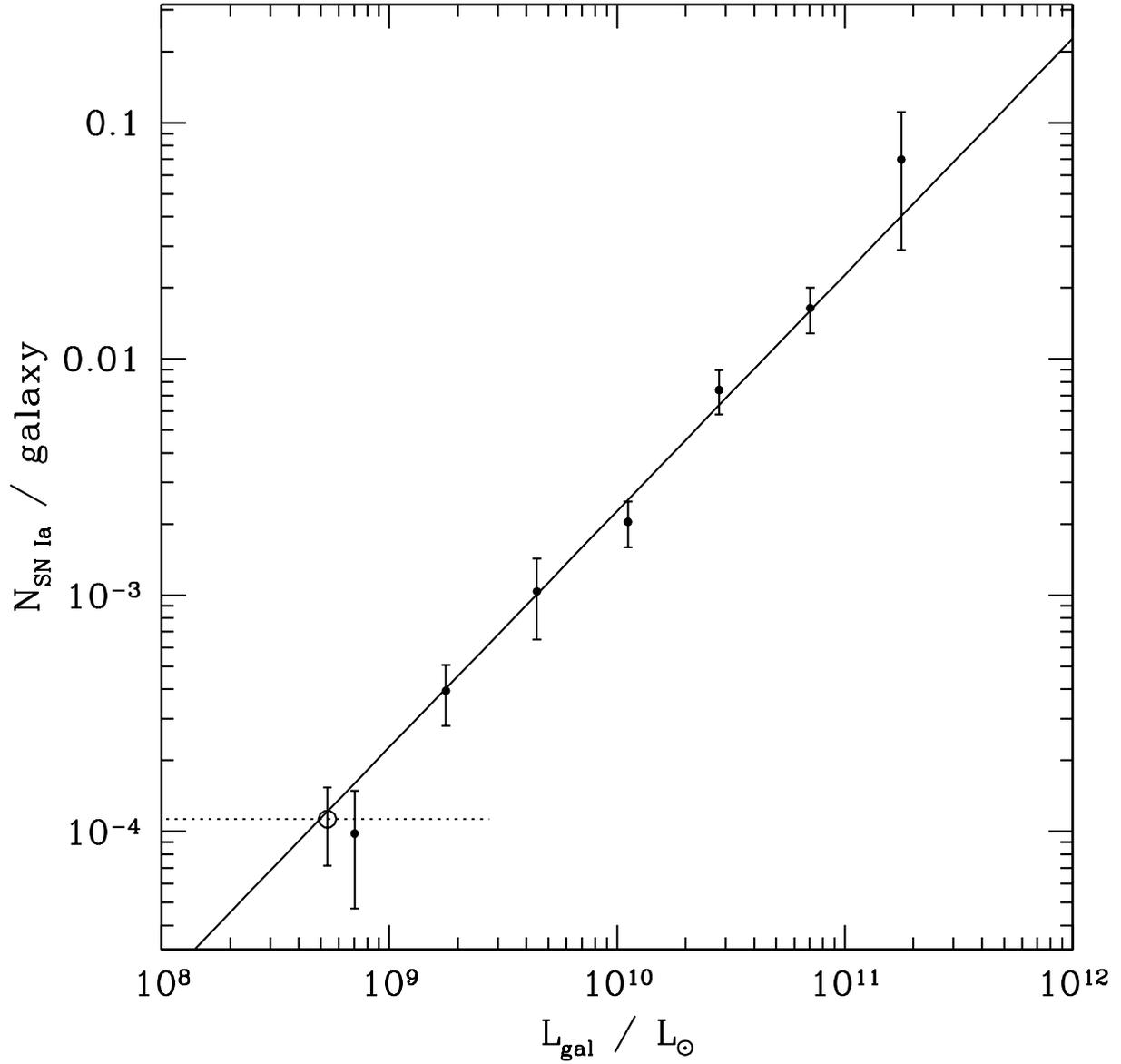}
\caption{SN Ia rate per galaxy as a function of $r$-band luminosity
of galaxies. The point denoted by the open symbol
stands for the hostless SNe Ia and the horizontal bar shows the
upper limit of luminosity of galaxies.
\label{fig:LumNum}}
\end{figure}

Figure \ref{fig:LumNum} shows that the SN rate per galaxy is
proportional to luminosity of host galaxies, as one expects from the
comparisons of the luminosity functions. The slope of the solid line
is fixed to the rate at 0.227 SNu(r) of Figures \ref{fig:hostlf_r}.
The data point at the bottom left indicated with the open circle is
for the `hostless SNe' where the horizontal error bars indicate only
the upper limit on host galaxy's luminosity. We conclude that our
hostless SNe Ia are consistent with them being occurred in low
luminosity galaxies beyond our detection limit but at the rate
proportional to luminosity of host galaxies.

\begin{figure}[ht]
\plotone{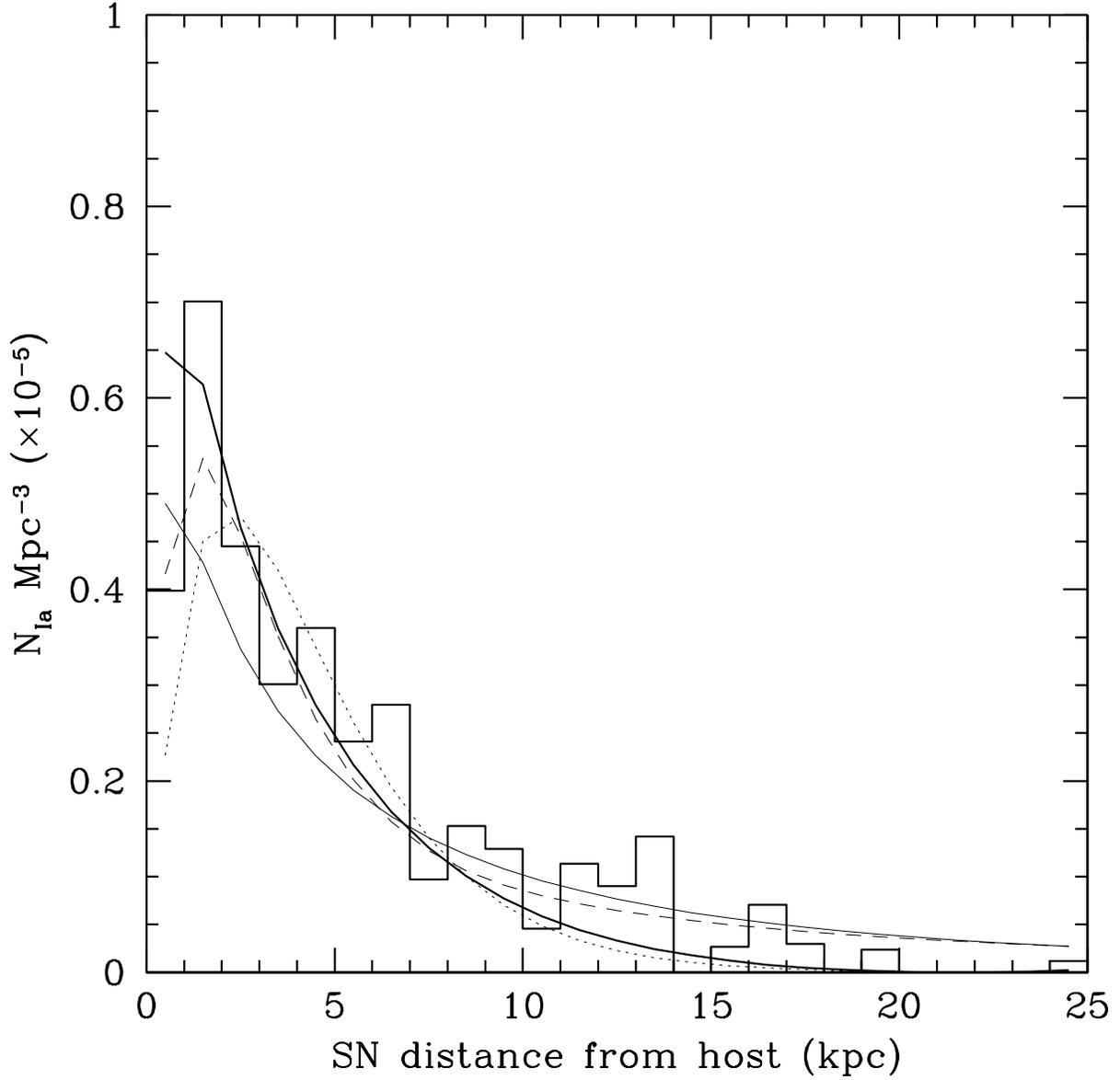}
\caption{Radial distribution of SNe Ia from the centres of their
host galaxies measured in physical distance scale. 
The thin solid, dotted and dashed
lines are best fit models with de Vaucouleurs, exponential
profiles and sum of these two profiles, respectively.
The thick solid line is the mean $r$-passband galaxy
light profile averaged over field galaxies.
\label{fig:dpos}}
\end{figure}

We present in Figure \ref{fig:dpos} the radial distance distribution
of SNe Ia measured from the centre of host galaxies measured in the
$r$-passband in physical distance units, where we do not correct
for the projection effect caused by inclination of host
galaxies. The relative distance between SN and the centre of host
galaxy is measured within the accuracy of 0.1 arcsec, which
corresponds to 0.4 kpc at $z=0.3$ \citep{Pier03}.  The ordinate
denotes numbers of the SN Ia per unit volume calculated by the formula
similar to eq. (\ref{eqn:analysis}).  This radial distribution is well
represented by the de Vaucouleurs profile with the half light radius of
$r_e = 5.7$ kpc as drawn by the thin solid curve in Figure \ref{fig:dpos}.
We also draw the exponential profile (dotted curve) with the half light
radius of $r_e=3.6$ kpc, which falls off faster than the distribution of
SNe Ia at an impact parameter larger than 10 kpc. The $\chi^2$ of the
best fit models, $\chi^2=8.2/11$ for the de Vaucouleurs profile and
$\chi^2=11.7/11$ for the exponential profile, where fitting was done
within 13 kpc using 13 data points, differ only a little but the
difference is more apparent in the tail.  When one uses the addition
of the de Vaucouleurs and the exponential profiles, the best fit model
shows bulge-to-disk luminosity ratio of 0.70, $r_e ({\rm deV}) = 9.1$
kpc and $r_e ({\rm exp}) = 2.4$ kpc with $\chi^2=5.0/9$. This fitted
profile is drawn by thin dashed curve.

Our result does not agree with that of 
\citet{Bartunov07} who claimed that SNe Ia in spiral galaxies are in a
lower rate in the central part compared to SNe Ia in elliptical galaxies
based on their supernova catalogue obtained by a compilation of SNe 
in the literature. Their earlier paper \citep{Bartunov92} claims that
the radial dependence of surface density of SNe Ia can be expressed
by the exponential profile. Our result does not agree with their 
conclusion, either. 

The thick solid curve in Figure \ref{fig:dpos} shows the empirical
mean $r$-passband light profile of field galaxies, which is
constructed from the aperture flux of galaxies at $z=0.025-0.030$ in
Catalog Archive Server of SDSS Data Release 6 using values of {\tt
profMean}.  The innermost bin may be somewhat affected by finite size
seeing. The normalisation is set to the same as Figure
\ref{fig:hostlf_r} while taking into account the luminosity evolution
of galaxies from $z=0.025-0.030$ to $z=0.2$ using the $Q$ parameter of
\citet{Blanton03} while fixing the shape of the light profile.  The light
profile of galaxies is consistent with the exponential disc plus de
Vaucouleurs spheroid model with the bulge-to-disk luminosity ratio of
0.56. We note that the radial distribution of SNe Ia, at least for the
bulk of SNe, is consistent with the light distribution
($\chi^2=11.2/13$), except for that at a large distance beyond 10 kpc
where we see some excess occurrence of SNe.  It is interesting to see
that some SNe Ia occur at a large distance where the galaxy
contributes little light: at $>10$kpc we expect from the global rate,
$1.31\times10^{-5}$ Mpc$^{-3}$yr$^{-1}$ SNe Ia with the light distribution of
the de Vaucouleurs
profile and $0.15\times10^{-5}$ Mpc$^{-3}$yr$^{-1}$ SNe Ia with the
exponential profile, which are compared with observed
$0.55\times10^{-5}$ Mpc$^{-3}$yr$^{-1}$ SNe Ia. The radial
distribution of SNe Ia also supports the proposition that the
occurrence of SNe Ia is primarily proportional to the luminosity of
host galaxies.

\begin{figure}[ht]
\plotone{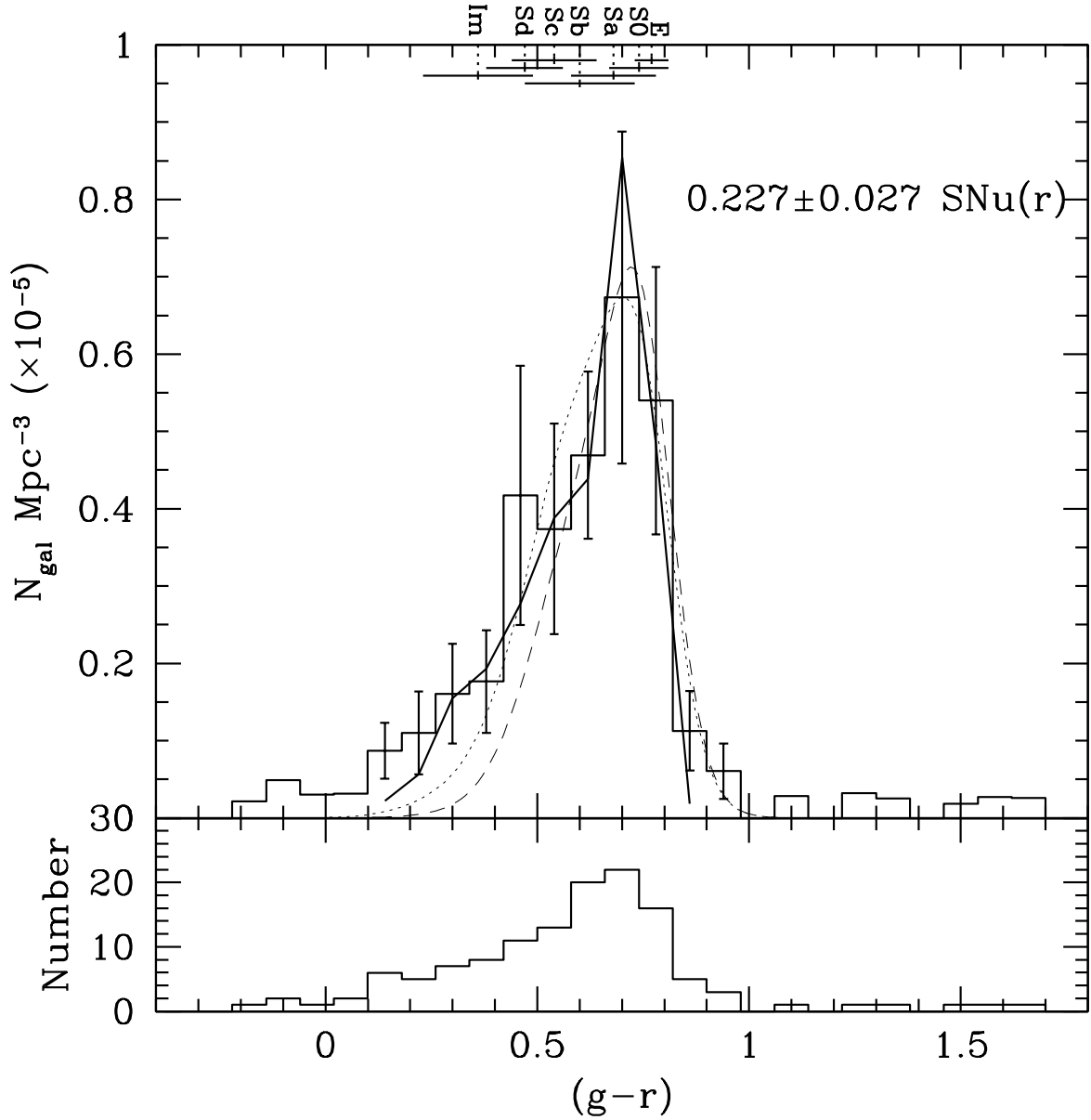}
\caption{$g-r$ colour function of SNe Ia host
galaxies with the lower panel indicating the number of contributing 
galaxies in each bin. 
Solid curve represents the luminosity weighted galaxy colour function
normalized in the way same as in Figure \ref{fig:hostlf_r}.
Morphological types of galaxies are indicated at the corresponding
colours according to \citet{Fukugita07}. Horizontal bars indicate 
the variance within the morphological types. Dotted and dashed lines are
numbers expected for SNe Ia based on the SNe Ia rate model of 
\citet{Mannucci05} and \citet{Sullivan06}, respectively.
\label{fig:hostcf}}
\end{figure}

We study the dependence of the SN rate on colour of host galaxies,
to examine whether the occurrence of SNe Ia would correlate with 
the star formation activity. 
In Figure \ref{fig:hostcf} the histogram represents the SN host galaxies and
the solid curve shows
the luminosity weighted colour function of field galaxies which is
calculated using bivariate function $\phi(M_r, g-r)$ \citep{Blanton01}, as
\begin{equation}
\Phi_L(g-r)d(g-r) = \int_{-24.25}^{-14.75} dM_r \phi(M_r, g-r) \times 10^{0.4(M_{\sun,r}-M_r)} d(g-r),
\end{equation}
where $\phi(M_r, g-r)$ is evaluated 
for $-24.25 < M_r < -14.75$ and $0.12 < g-r < 0.88$ on the $20\times20$
grid. The abscissa is rest-frame colour after the K-correction.
The normalisation is taken to be the same as that in Figure
\ref{fig:hostlf_r}. Morphological types of galaxies are indicated with
the corresponding colours according to \citet{Fukugita07} with the
error bar representing the dispersion of colours in the morphologically 
classified sample. This figure shows that
the colour distribution of SNe Ia host galaxies traces well that
of field galaxies, and we do not see
any particular excess of SNe Ia host galaxies for bluer, late type or
irregular galaxies, beyond one sigma level. 
In this comparison we do not take into account the evolution of galaxy
colour to $z = 0.2$.

In particular we may be interested in the difference of the SN rate 
between the elliptical galaxies
and other galaxies. If we select galaxies with colour $0.73<g-r<0.81$ 
corresponding to elliptical galaxies (with contaminations from S0 and some Sa
galaxies) we obtain the SN rate
$0.194\pm0.062$ SNu(r) which is compared with $0.255\pm0.035$
for galaxies with $g-r<0.73$ (Sa galaxies or later). We observe
a 31$\pm35$\% enhancement in the SN rate relative to galaxy luminosity
in late type galaxies compared to that in elliptical galaxies, 
though the effect is 
only at one sigma. We do not see a particular enhancement of SN rate
per luminosity, however, among late type galaxies, where the star formation
rate increases towards later type morphologies \citep{Nakamura04}.

$u-g$ colour is more sensitive to the star formation activity
than $g-r$, but $u$-band brightness is too faint to us for most host
galaxies: SN host galaxies brighter than 21 mag are less than 40\% for
our magnitude limit supernova sample, and we are not able to draw a
meaningful conclusion from the $u-g$ colour distribution.
The SN host galaxies are also too faint for SDSS spectroscopy
\citep{Strauss02},
and we cannot estimate the star formation rate directly,
unless we resort to population synthesis colours, which are not well
constrained without $u$ colour. Therefore, we do not
find clear evidence that points to the correlation of the SN rate
with the star formation activity. 

\citet{Mannucci05} and \citet{Sullivan06} give models of SN Ia rates with 
the two components of progenitors, 
one explosions of old binary stars with a large delay time
from the formation and the other explosions with a short delay time
after the system formed. \citet{Mannucci05} give 
SNe Ia rate that depends on the morphological type of host
galaxies in units of SNu($K$). We may use their rate to estimate the
number of SNe Ia for each morphological type using morphological-type
dependent luminosity density of field galaxies, as given by
\citet{Nakamura03} with the aid of the conversion from the $r$ to
the $K$-passband \citep{Nagamine06}. This model gives 
$r_V=3.0\times10^{-5}\;{\rm Mpc}^{-3}\;{\rm yr}^{-1}$,
where 31\% are the prompt component, which is consistent with the
difference in SN Ia rate relative to galaxy luminosity between 
elliptical galaxies and spiral galaxies.
The total rate also agrees with
our rate in equation (\ref{eq:volsnr}).
The detailed numbers are presented in Table \ref{tbl:rate}.

\citet{Sullivan06} modelled SN Ia rate as a function of stellar
masses and mean SFRs of host galaxies. We use the morphology dependent
luminosity density of \citet{Nakamura03} and the morphologically dependent
star formation rate calculated from the star formation rate of the bugle
and disk components given in \citet{Nagamine06}.
The predicted SN Ia rate is 
$r_V=2.6\times10^{-5}\;{\rm Mpc}^{-3}\;{\rm yr}^{-1}$,
of which 13\% are prompt: see also Table \ref{tbl:rate}. 
The predicted numbers of SNe Ia are plotted in Figure
\ref{fig:hostcf} as dotted (Mannucci et al.'s model) and dashed
(Sullivan et al.'s model) curves taking the colour distribution of
each morphological type as a Gaussian with mean and dispersion given by
\citet{Fukugita07}.  The $\chi^2$ for the two models are 
15.1 and 23.6 for 11 degrees of freedom using data for
$0.1<g-r<1.0$. 

The currently available two component models are marginal to represent
our data, but
are consistent with the observed
number of SNe Ia, as shown by the histogram in Figure \ref{fig:hostcf}
allowing for large uncertainties.
The prompt component is rather minor and not very manifest in the integrated
rates for low $z$ SNe, as in our analysis. The model prediction 
for Im galaxies is low, which is due
to the low luminosity density of Im galaxies in the SDSS morphologically
classified sample. In \citet{Nakamura03} Im galaxies contribute only
$\sim1\%$ of the total luminosity density, which may be due to the
selection effect that disfavours Im galaxies. This contrasts to the
indication from the colour function of \citet{Blanton01}, which
suggests that galaxies bluer than $(g-r)<0.42$ contribute by $\sim15\%$
of total luminosity.  If the luminosity density of Im galaxies is
this large, however, the number of SNe Ia in Im galaxies would be 10
times more that would largely overshoot our observed numbers
of SNe. To identify the prompt component and test the two component
hypothesis, we need a sample where the prompt component is dominant, 
or the sample with which both star formation rate and SN rate can 
more accurately be measured.

\begin{figure}[ht]
\plotone{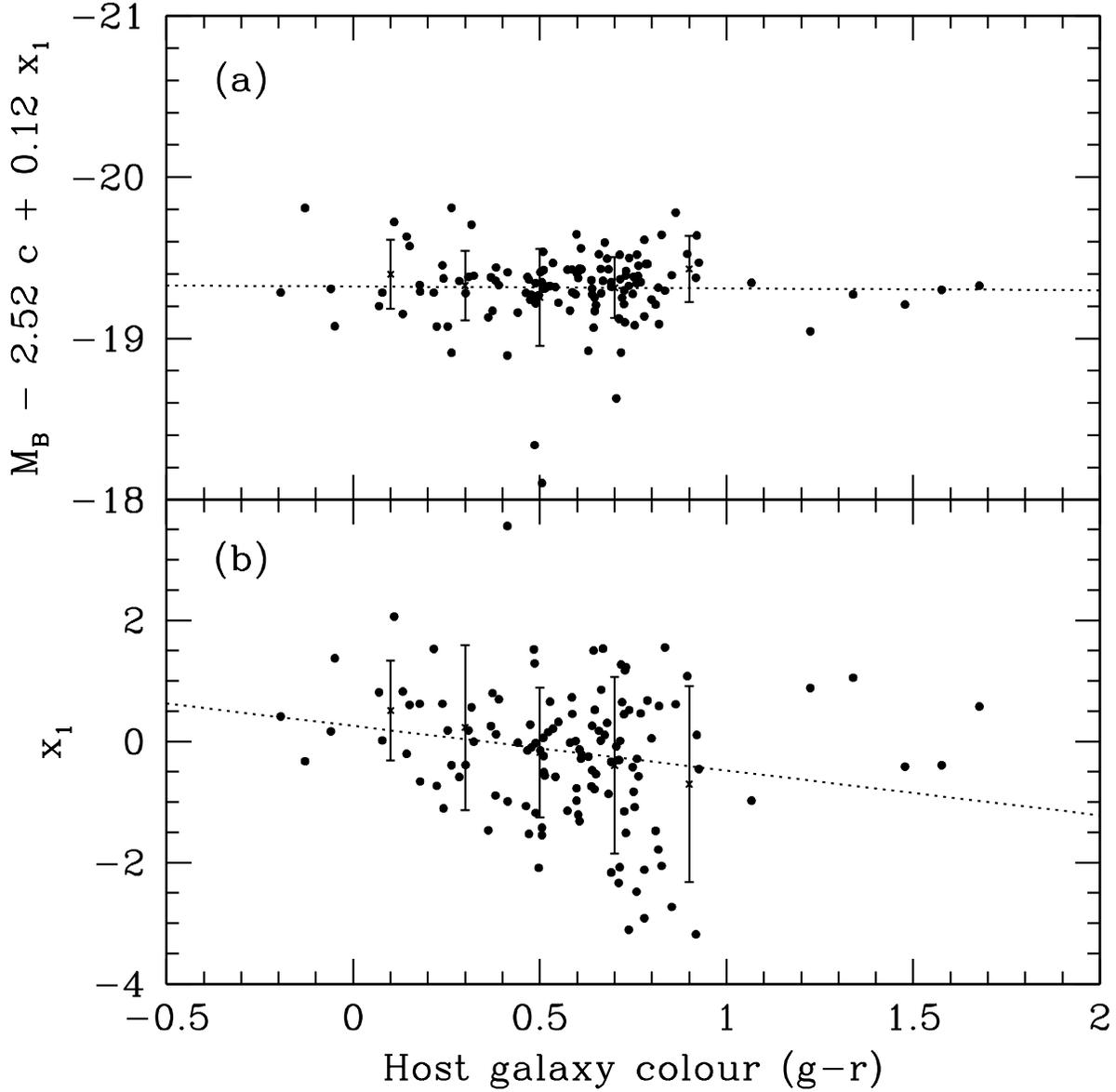}
\caption{(a) Absolute brightness of SNe Ia corrected for the colour excess
using $\beta = 4.1$ as a function of $(g-r)$ colour of host galaxies.
The mean and dispersion of the points are also plotted.
(b) Shape parameter $x_1$ that depends on colour $(g-r)$ of host galaxies.
Dotted lines show least square fits to the data, 
$M_B-4.1c=-19.32+0.01 (g-r)$ for (a), and $x_1=0.26-0.74(g-r)$ for (b),
respectively. 
\label{fig:snmhostc}}
\end{figure}

Our sample allows us to study if the properties of SNe Ia would depend
on the property of host galaxies.  It has occasionally been claimed
that bright SNe Ia appear more often in late type galaxies, while SNe
Ia in early type galaxies are subluminous \citep{Filippenko89,
  DellaValle92, Reindl05, Sullivan06}.  This has been counted as one
of the reasons that SN Ia calibration of nearby galaxy's distance
led to a smaller value of the Hubble constant in the past. Figure
\ref{fig:snmhostc}(a) plots the maximum brightness of SNe Ia in our
sample as a function of $g-r$ colour of host galaxies: mean brightness
of SNe Ia does not change across colours of galaxies from E to Im.
In particular the plot does
not indicate any evidence that SNe Ia in bluer galaxies are
systematically brighter or those in red galaxies are fainter.  Both mean and
dispersion of luminosity are nearly constant with
respect to galaxy colour as seen in the figure.

Figure \ref{fig:snmhostc}(b) shows the plot of the $x_1$ parameter,
which represents the decline rate of SNe Ia brightness, 
against $g-r$ colour of host galaxies. 
The upper values of $x_1$ does not change with respect to colour from
E to Im galaxies, but the lower value shows the trend that it decreases
from $-0.5$ (or $\Delta m_{15}\simeq 1.2$) for Im galaxies to $-3$
($\Delta m_{15}\simeq 1.7$) for E and S0 galaxies. It is noted that
large negative values of $x_1$ are seen only in early-type galaxies.
This causes some dependence of $x_1$ on colour of host galaxies
as seen in the figure. The slope, however, is small enough to cause
any effect on brightness of SNe Ia through the correction of
$0.12x_1$.

\begin{figure}[ht]
\plotone{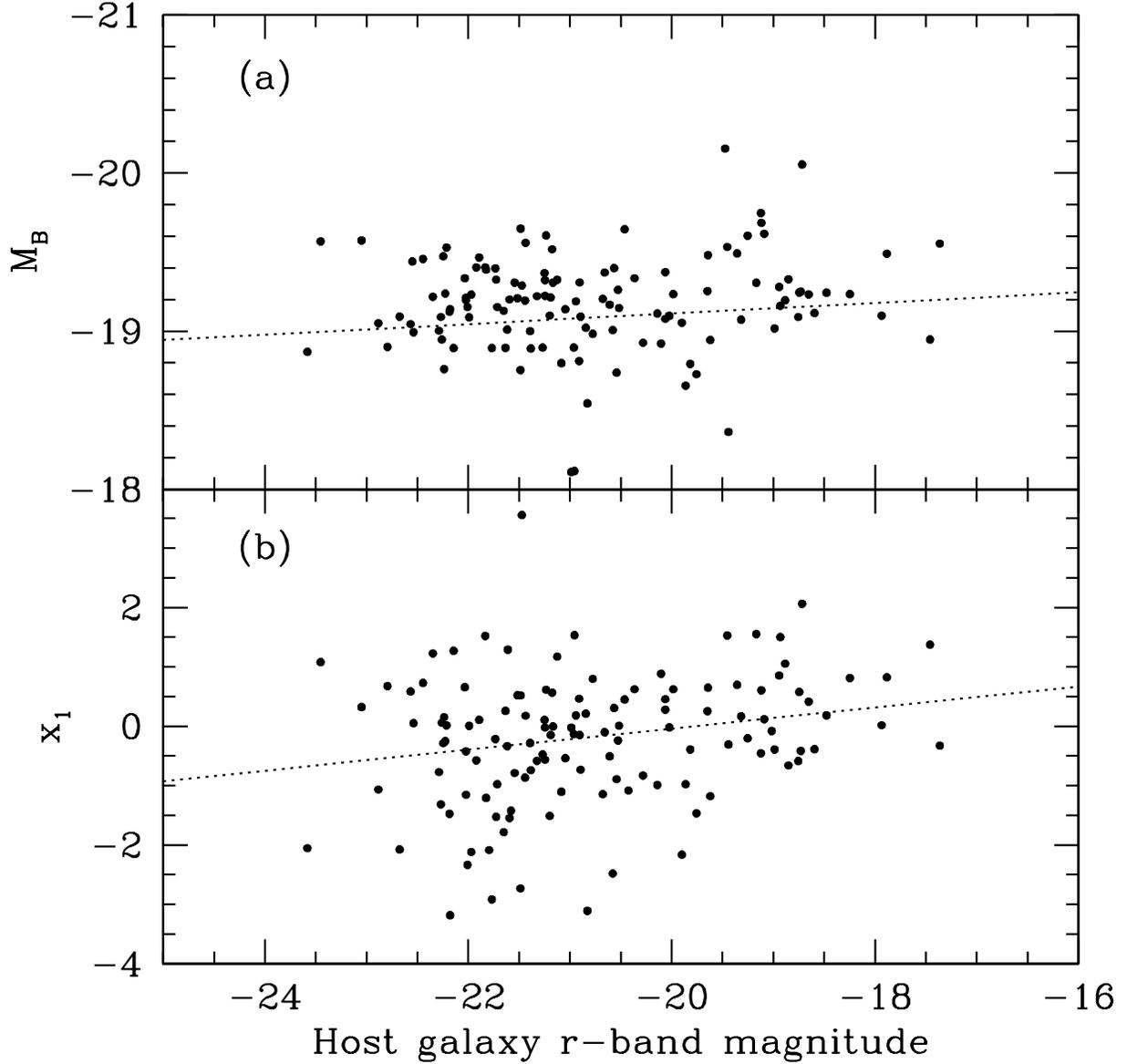}
\caption{(a) Absolute brightness of SNe Ia corrected for the colour excess
using $\beta = 4.1$ and (b) shape parameter $x_1$ as a function of
$r$-band brightness of host galaxies.
Dotted lines show least square fits to the data,
$M_B-4.1c=-19.78-0.034M_r$ for (a), and $x_1=3.50+0.18M_r$ for (b),
respectively. 
\label{fig:snmhostm}}
\end{figure}

Similarly Figure \ref{fig:snmhostm} shows maximum brightness of SNe Ia on 
luminosity of host galaxies, showing that the former does not depend on
luminosity of the host.
Note that we expect that average metallicity of the host galaxy
changes by 0.59 dex in this luminosity range \citep{Tremonti04}.
The $x_1$ parameter
changes slightly as luminosity of host galaxies changes.

\begin{figure}[ht]
\plotone{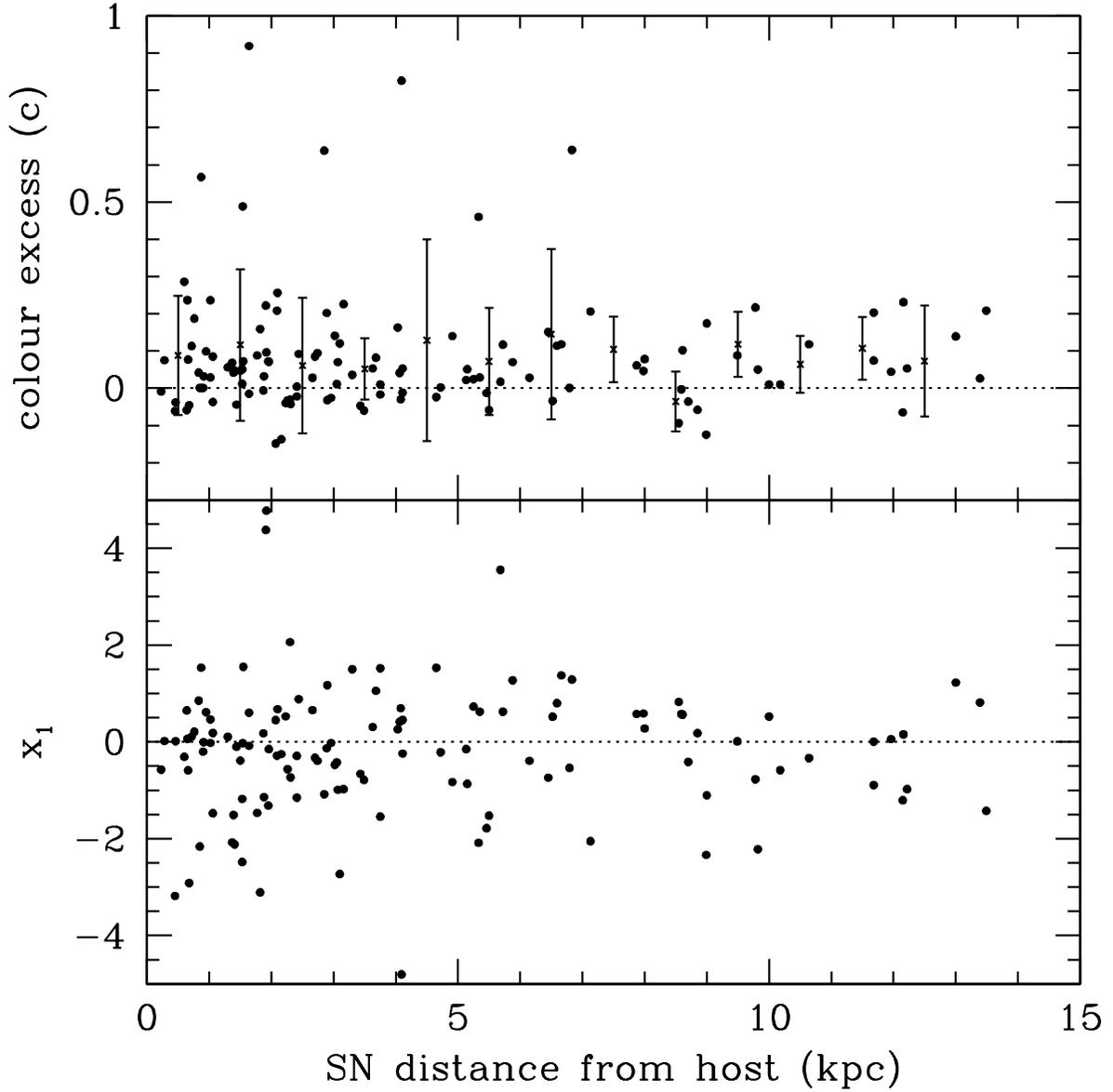}
\caption{Colour excess and the $x_1$ parameters plotted against 
the physical distance of SNe Ia
from the centre of host galaxies. Dotted lines indicate
the zero level. In upper panel the mean and dispersion in respective
bins are also plotted.
\label{fig:dpos_color}}
\end{figure}

We also examine the dependence of the $c$ parameter and the $x_1$
parameter on the distance from the centre of galaxies as shown in
Figure \ref{fig:dpos_color}. The colour excess $c=E(B-V)$ stays nearly
at $\approx$0.08 with the dispersion of 0.07 and does not show a
systematic change from 1 kpc to 15 kpc.  In particular, we observe
that the colour excess does not decrease as we go farther away from
the centre of galaxies, where stars, and hence dust, are expected to
decrease. 
The trend we see here does not change if we
limit the SNe to $z<0.2$ with which SNe Ia that would receive
reasonably large extinction are included in the sample.  This finding
suggests us to interpret that the observed reddening is mostly
associated with individual supernovae, either extinction from
supernova itself and/or circum-supernova dust or intrinsic colour
variation of the supernova, rather than interstellar dust in host
galaxies. This would also give us an upper limit on the model as to
the amount of dust produced around SNe Ia. We observe only 7 SNe Ia
which shows colour excess more than $E(B-V)\geq 0.3$ mag among our 137
SN Ia sample.  In addition, we do not observe a systematic dependence
of the $x_1$ parameter on the distance from the centre of galaxies;
see Figure \ref{fig:dpos_color}(b).

\section{Conclusion}

The SNe Ia sample acquired in the SDSS II, containing 137 low redshift
SNe from $z=0.05$ to 0.3, indicates that the occurrence of SNe Ia is
primarily proportional to luminosity of galaxies. The LF of SN Ia host
galaxies matches very well with that of field galaxies multiplied by
luminosity, and colour of SN Ia host galaxies does not differ from
that of field galaxies.  Our low redshift sample does not indicate an
active signature that the occurrence of SNe Ia follows star formation
activity, except that possible enhancement in the SN Ia rate is
noted in late type galaxies ($\lesssim 31\pm35$\%) compared with the
rate in elliptical galaxies.  Our low redshift sample is compatible
with the two component model where the effect of the prompt component
is modest (10$-$30\% of the total rate), as in the current models.  
We are not able to
differentiate SN rates among late type galaxies.  Our sample contains
8 SNe Ia, whose host galaxies were not identified.  It is shown,
however, that they are consistent with them occurred in low luminous
galaxies beyond the survey limit for galaxies. Luminosity of SNe Ia
does not appear to depend upon luminosity or colour of host
galaxies.

The luminosity function of SNe Ia is Gaussian with $M_B=-19.42$ and
$\sigma=0.24$ mag (FWHM is a factor 1.4 in luminosity), if the colour
variation is interpreted as reddening obeying the extinction law of
the Milky Way with the standard value $R_V=3.1$.  This Gaussian
distribution is further tightened and $\sigma=0.14$ mag if the
extinction to reddening ratio $R_V$ is reduced to $\approx 2$ and if
the correlation is taken into account between maximum brightness and
the decline rate.  Reddening of SN Ia, inferred from the colour
variation, is 0.1 mag in $A_V$, which is lower approximately by a
factor of 5 than that for star forming field galaxies measured from
H$\alpha$ and H$\beta$ emission.

We also showed that the distribution of the SN Ia occurrence, that
extends to a few tens of kpc, agrees with the general light
distribution of galaxies that have bulges with the de Vaucouleurs type
profiles.  The variation of colours of SN Ia is constant and does not
depend on the distance from the centre of galaxies unlike what is
expected for the supernova that would happen in the galactic disc. The
colour variation of SNe Ia may likely be ascribed to intrinsic colour
variation or immediate neighbourhood of supernovae rather than
extinction in host galaxies. The total column density of dust
should not give extinction more than 0.2 mag in the $V$ passband.

\acknowledgements
We should like to thank the review group of the SDSS-II Supernova
Survey Project (Josh Frieman, Bob Nichol, Saurabh Jha, Benjamin Dilday) and
Don Schneider and Richard Kessler  
for reviewing the manuscript and useful comments
improving the manuscript.  
NY acknowledges support by the JSPS core-to-core program ``International
Research Network for Dark Energy'' and the JSPS-USA bilateral programme,
MF is supported by Grant-in-Aid of the Ministry of Education in Japan,
and Ambrose Monell Foundation at Princeton. 

Funding for the SDSS and SDSS-II has been provided by the Alfred
P. Sloan Foundation, the Participating Institutions, the National
Science Foundation, the U.S. Department of Energy, the National
Aeronautics and Space Administration, the Japanese Monbukagakusho, the
Max Planck Society, and the Higher Education Funding Council for
England. The SDSS Web Site is {\tt http://www.sdss.org/}.

The SDSS is managed by the Astrophysical Research Consortium for the
Participating Institutions. The Participating Institutions are the
American Museum of Natural History, Astrophysical Institute Potsdam,
University of Basel, University of Cambridge, Case Western Reserve
University, University of Chicago, Drexel University, Fermilab, the
Institute for Advanced Study, the Japan Participation Group, Johns
Hopkins University, the Joint Institute for Nuclear Astrophysics, the
Kavli Institute for Particle Astrophysics and Cosmology, the Korean
Scientist Group, the Chinese Academy of Sciences (LAMOST), Los Alamos
National Laboratory, the Max-Planck-Institute for Astronomy (MPIA),
the Max-Planck-Institute for Astrophysics (MPA), New Mexico State
University, Ohio State University, University of Pittsburgh,
University of Portsmouth, Princeton University, the United States
Naval Observatory, and the University of Washington.

\appendix
\section{Verification of the analysis method}

In order to examine the validity of our analysis, we have made
simulations for our observation creating a set of light curves
using SALT2 model in combination with SNANA. The distribution
of $x_1$ parameter is assumed to be a Gaussian with the mean 0 and
the dispersion 0.90. The distribution of $c$ is a
Gaussian with the mean 0.063 and the dispersion 0.11.
We assume the relation involving absolute magnitude $M$, $x_1$ and $c$, 
\begin{equation}
M = M_0 - \alpha x_1 + \beta c
\end{equation}
with $\alpha = 0.10$ and $\beta = 2.45$.
The intrinsic dispersion of 0.11 mag was applied 
for every passbands and every epochs.
The epochs of observation is fixed to actual date and
the signal to noise ratio of each observing point was calculated by
using real observational condition of our observation including
seeing and sky brightness. The detection efficiency searching for
variable objects is also included in our simulation, and the 
SN Ia rate of $r_{\rm SNe Ia} = 2.2 \times 10^{-5}\;(1+z)^{1.5}\;({\rm Mpc}^{-3}yr^{-1})$ is assumed.
The cut based on the quality of light curves as described in the text
is also applied to the simulation, and the
spectroscopic incompleteness of eq. (\ref{eqn:completeness}) is
applied with $z_c = 0.15$ and $\Delta z = 0.3$.
The simulation generated 210 SNe
light curves.
We make 50 sets of simulated light curves and
applied the method the same as that to the observed dataset. The incompleteness
correction $\epsilon(z)$ in eq. (\ref{eqn:analysis}) is calculated 
separately for each simulation set.

\begin{figure}[ht]
\plotone{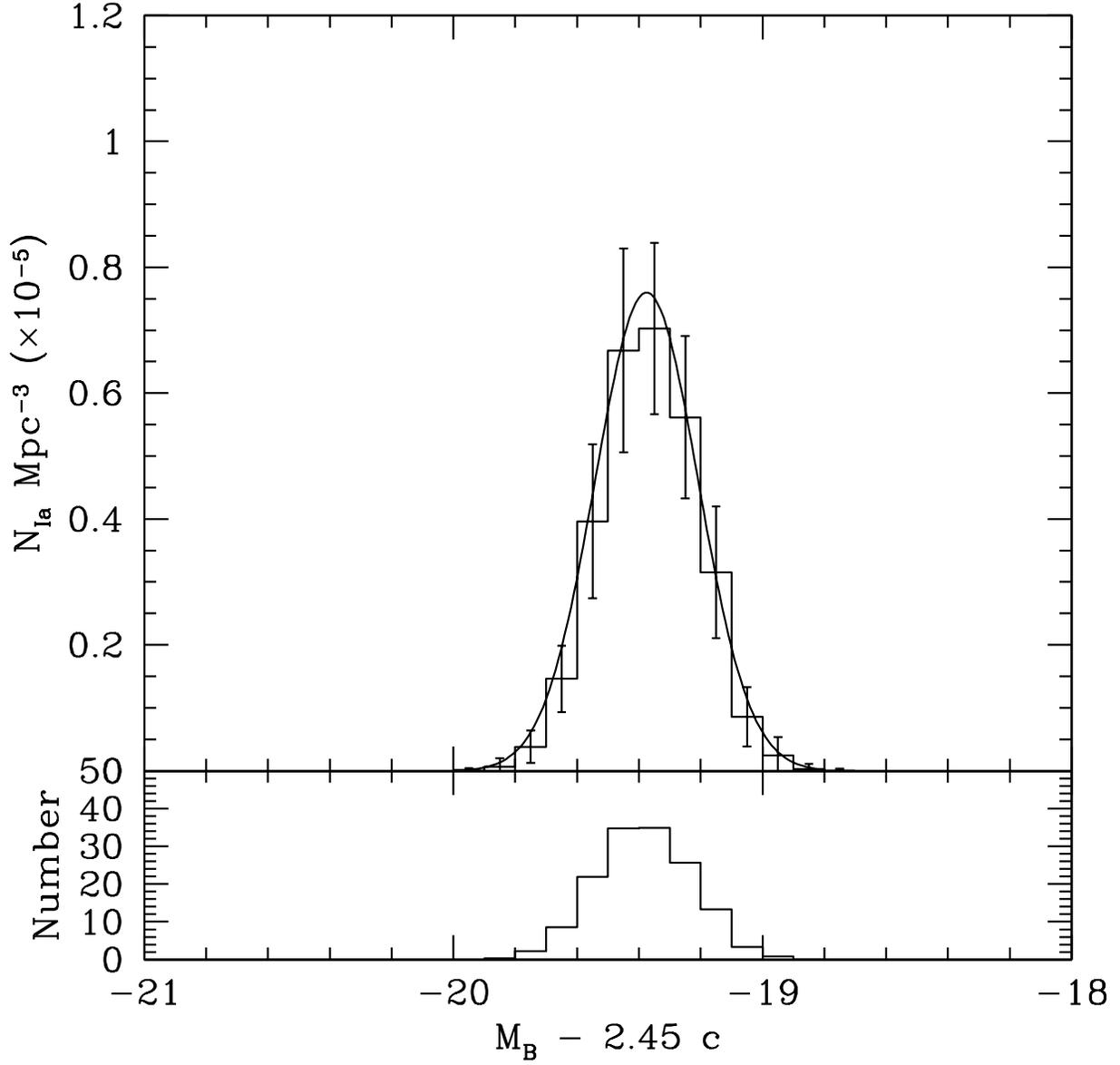}
\caption{Comparison of the input SN LF (solid curve) and output SN LF (histogram)
from simulated dataset assuming the completeness as a function of redshift.
Histogram passes through the mean and error bars show 
the dispersion in each bin calculated from 50 simulation dataset.
\label{fig:snlf_simz}}
\end{figure}

Figure \ref{fig:snlf_simz} shows the comparison of the input and the 
output SN LF after correcting for the effect of $c$ parameter. 
Solid curve represents the expected LF taking into account the error of
the determination of $m_B$ and $c$ from light curve fits. 
Histograms and error bars are the mean and dispersion in each magnitude bin
calculated from the 50 simulated light curve sets. 
A good agreement is seen between solid curves and histograms.

\begin{figure}[ht]
\plotone{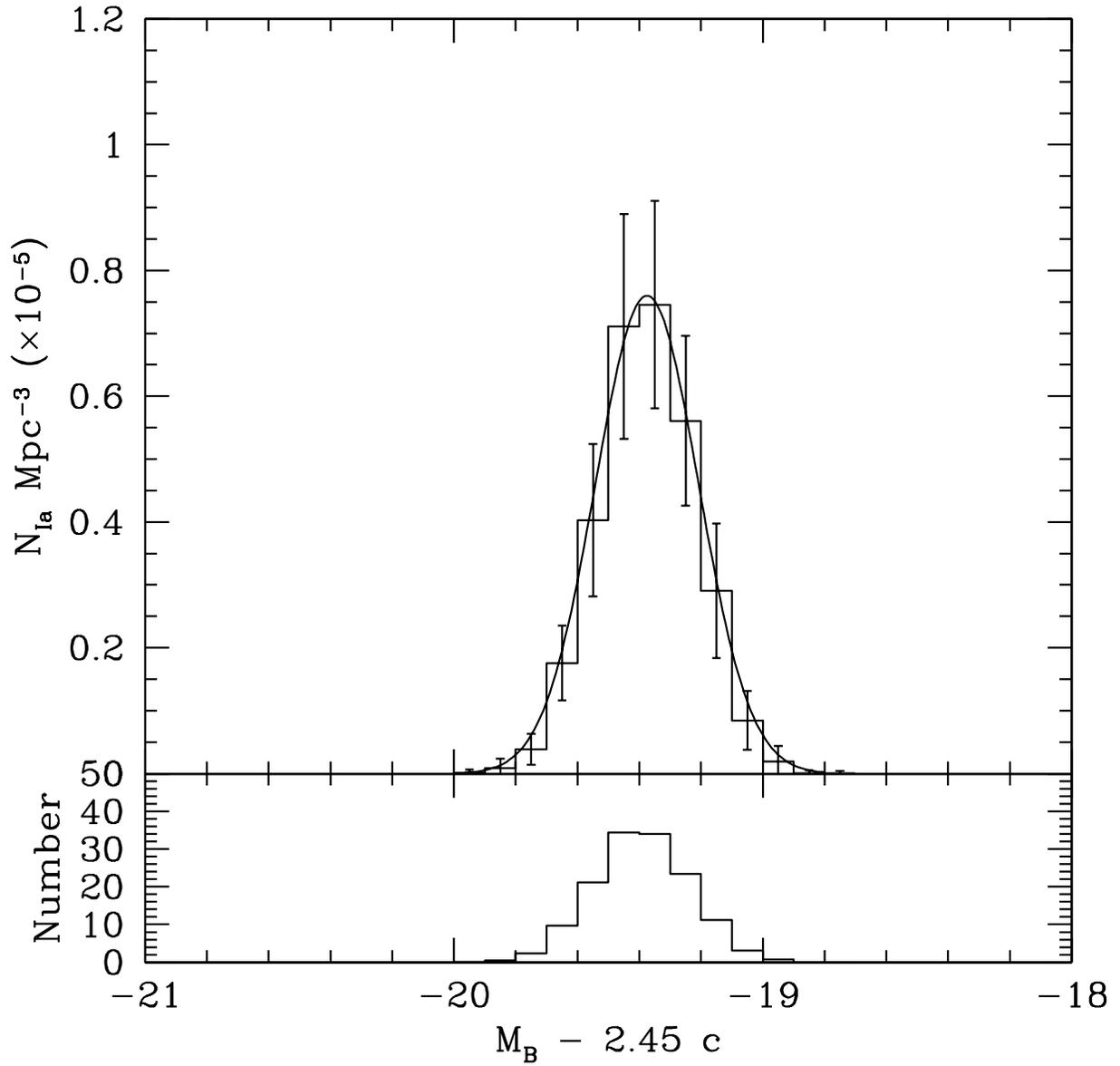}
\caption{Same as Fig. \ref{fig:snlf_simz} but completeness is
assumed to be a function of apparent peak $r$-band magnitude.
\label{fig:snlf_simm}}
\end{figure}

We also make a simulation incorporating the completeness as a function of
apparent peak $r$-passband brightness of SNe like
$\epsilon(m) = 1.0 - (r_{\rm max}-19.75)/3.15$ to see the effect when the
completeness is not a function of redshift but of apparent peak magnitude.
The completeness $\epsilon(z)$ in eq. (\ref{eqn:analysis})
is estimated as a function of redshift. The result is shown in Fig. 
\ref{fig:snlf_simm}. We see again a good agreement between the input
and output SNe. This is due to the fact that redshift and apparent peak 
magnitude are well correlated.

Good agreements are seen between input and output SN LF's for both cases
where incompleteness is a function of redshift and of apparent peak
$r$-passband magnitude. The analysis adopted in this paper does not
suffer from particular systematic effects while dealing with
apparently a dataset with sample incompleteness.
The volumetric SN rate is recovered to be $(2.95\pm0.42)\times10^{-5}$
and $(3.04\pm0.51)\times10^{-5}$ depending on whether incompleteness
is taken to be a function of redshift or apparent magnitude, respectively.
These numbers are compared with
the input value of $3.18\times10^{-5}\;({\rm Mpc}^{-3}\;yr^{-1})$.
The volumetric SN rate is properly recovered within the error.

\begin{figure}[ht]
\plotone{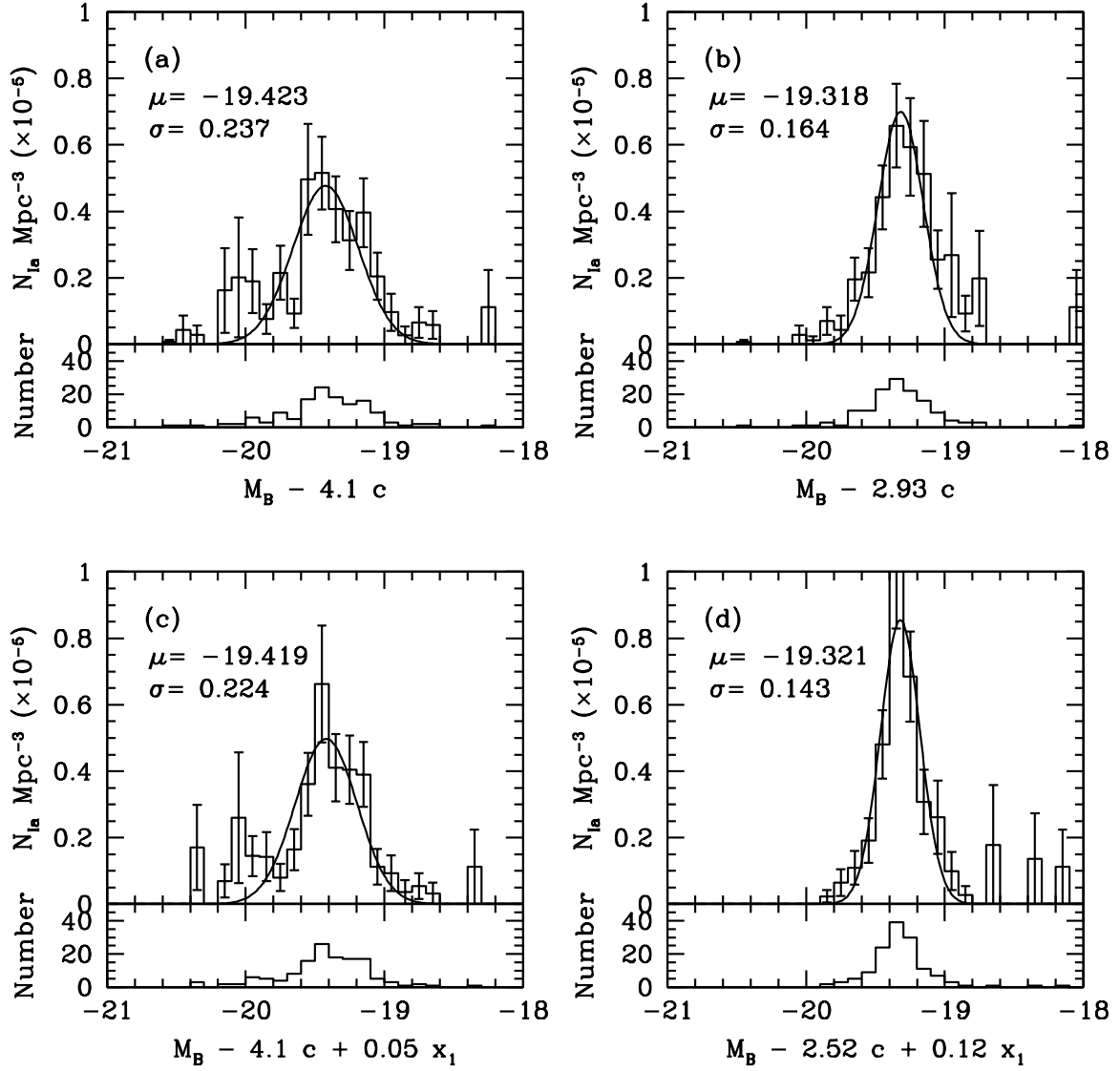}
\caption{Luminosity function of SNe Ia in the $B$-band. $B$-band
brightness is corrected for the colour variation as $M_B-\beta c$
($\beta=R_V+1$)
with (a) $\beta = 4.1$ and (b) $\beta=2.92$ which minimises the width
of fitted Gaussian. The parameters of the 
Gaussian are shown 
in each panel. In panels (c) and (d) a correction for
light curve shape $x_1$ is also taken into account with the parameter
to minimise the Gaussian widths.
\label{fig:snlf}}
\end{figure}

\begin{figure}[ht]
\plottwo{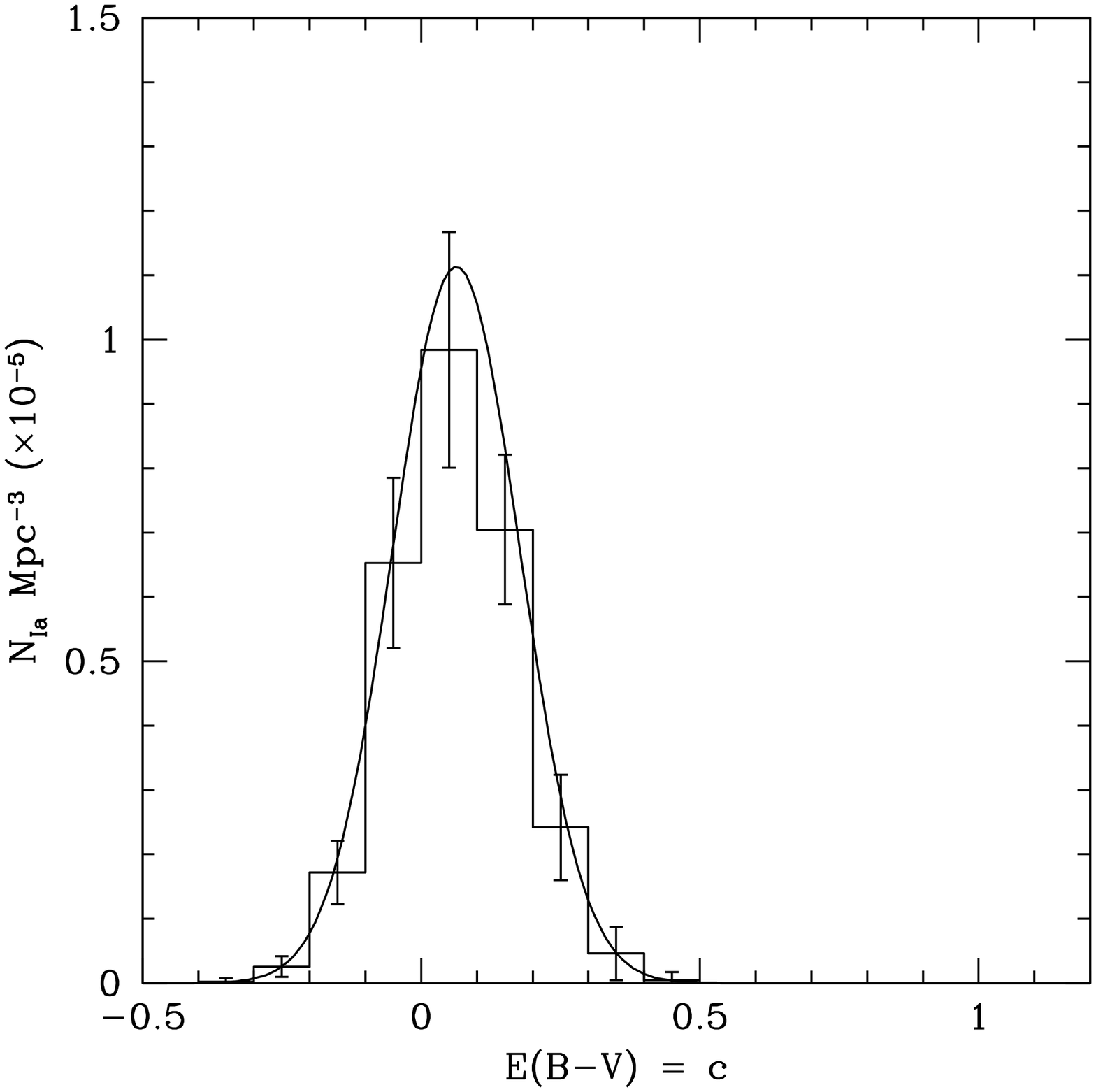}{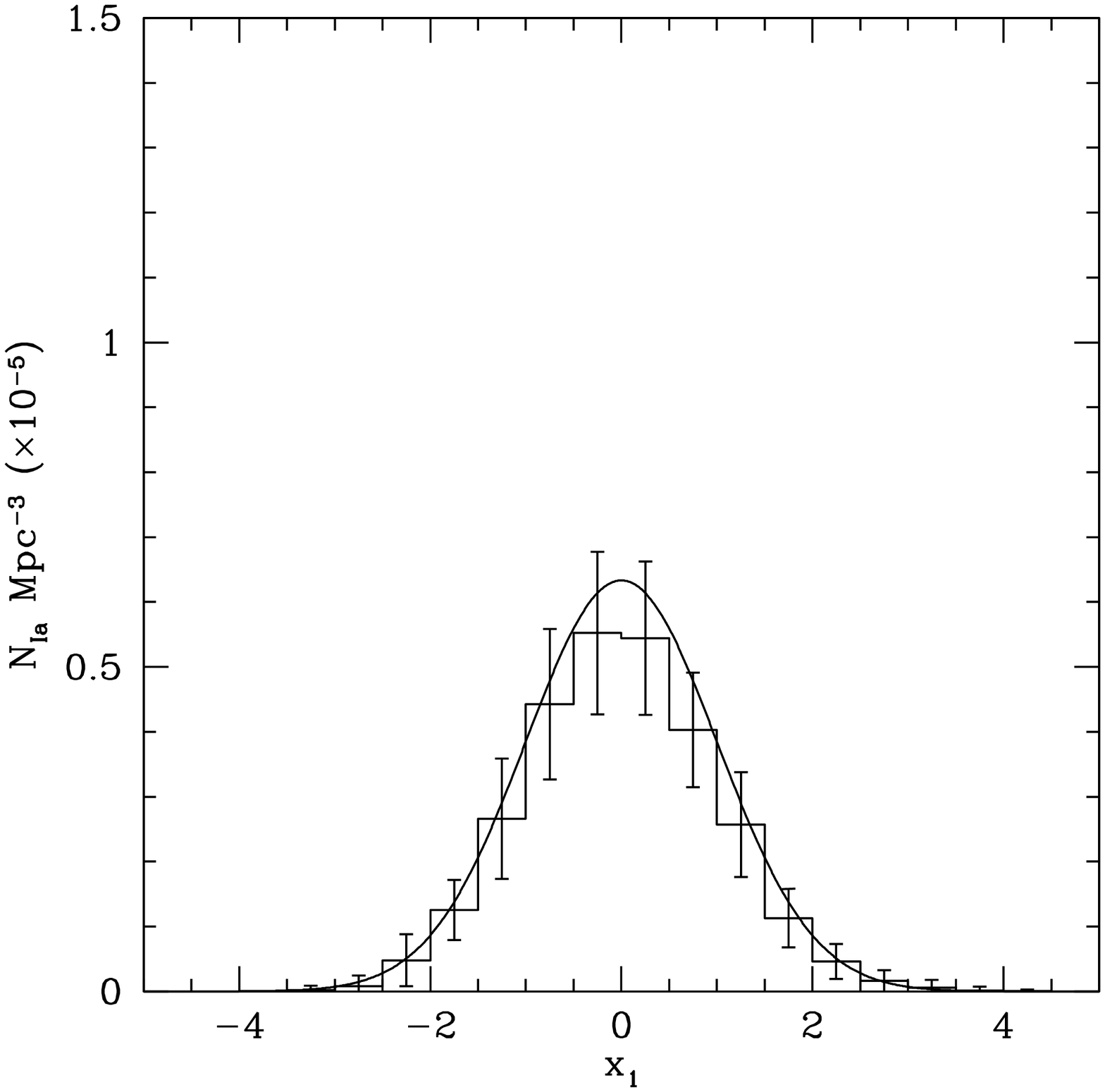}
\caption{Same as Fig. \ref{fig:snlf_simz} but for the distribution
of the $E(B-V) = c$ parameter (left) and the $x_1$ parameter (right).
\label{fig:ebvhist_sim}}
\end{figure}

\begin{figure}[ht]
\plotone{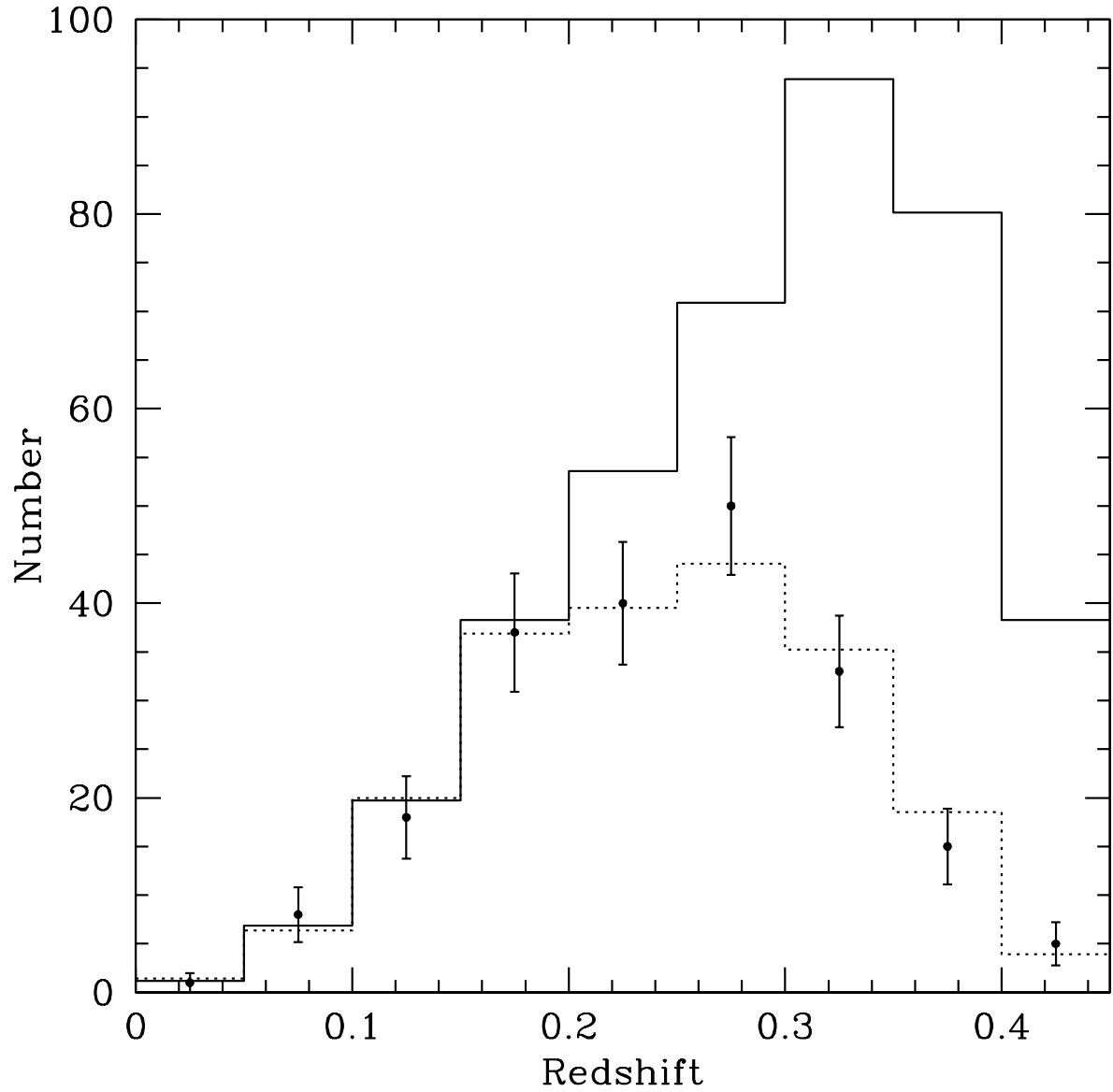}
\caption{Simulated redshift distributions before (solid) and after (dotted)
spectroscopic incompleteness is applied. Points with error bars represent
observed SNe sample.
\label{fig:zhist_sim2}}
\end{figure}

We also examine the distribution of the $c$ and $x_1$ parameters of SALT2
light curve fit in Fig. \ref{fig:ebvhist_sim}. SNe with high $c$ value
or low $x_1$ value will be faint and such SNe may suffer from 
incompleteness. The analysis is similar to that of the LF. In
eq. \ref{eqn:analysis}, we replace absolute magnitude $M$ with the
colour parameter $c$ or the shape parameter $x_1$. We do not see any
systematic effect for the recovered distribution. 
We change the assumed distributions of the $c$ and $x_1$ parameters
and the relation among $M$, $x_1$ and $c$ in a way they are consistent
with the observed SNe, and calculate the completeness of the observed
sample. This procedure is iterated till the convergence (one or two
iterations are sufficient). The final values are
$x_1=0\pm0.96$, $c=0.039\pm0.093$ and $M=M_0-0.12x_1+2.52c$, where
the last relation agrees with the one obtained by minimising the
width of the LF of SNe Ia as seen in Fig. \ref{fig:snlf}(d).  With these
procedures the redshift distribution of the observed SNe is
reproduced as shown in Figure \ref{fig:zhist_sim2}.

We conclude that the method we used in this paper works well even for
the data with incompleteness after the proper account taken for
incompleteness. We do not expect any specific biases in the analysis.

\begin{deluxetable}{lr}
\tablecolumns{2}
\tablewidth{0pc}
\tablecaption{Sample selection summary\label{tbl:sample}}
\tablehead{
\colhead{Criteria} & \colhead{Number}}
\startdata
total       & 314 \\
~~~~spectroscopic/confm'd & 130 \\
~~~~spectroscopic/probable & 16 \\
~~~~light curve selection & 168 \\
secure light curve & 222 \\
good sky area \& date & 207 \\
$r_{\rm lim} < 21.5$ & 137 \\
~~~~no host galaxies & 8 \\
\enddata
\end{deluxetable}

\begin{deluxetable}{ccccc}
\tablecolumns{3}
\tablewidth{0pc}
\tablecaption{Hostless SNe Ia\label{tbl:hostless}}
\tablehead{
\colhead{SDSS ID} & \colhead{IAU Name} & \colhead{Redshift} & \colhead{Luminosity Limit ($L^*_r$)} & \colhead{Dist. of 10\arcsec (kpc)}}
\startdata
2943 & 2005go & 0.2659 & 0.055 & 40.7 \\
5994 & 2005ht & 0.1885 & 0.025 & 31.2 \\
6780 & 2005iz & 0.2046 & 0.029 & 33.1 \\
6924 & 2005ja & 0.3286 & 0.089 & 47.2 \\
6933 & 2005jc & 0.2137 & 0.033 & 34.5 \\
7475 & 2005jn & 0.3188 & 0.085 & 46.7 \\
7335 & 2005kn & 0.1975 & 0.028 & 32.6 \\
3565 &        & 0.2885 & 0.067 & 43.4 \\
 & & & \\
8030\tablenotemark{a} & 2005jv & 0.4226 & 0.161 & 55.5 \\
\enddata
\tablenotetext{a}{Outside the magnitude limited sample}
\end{deluxetable}

\begin{deluxetable}{lcc}
\tablecolumns{2}
\tablewidth{0pc}
\tablecaption{SNe Ia Rate Measurements\label{tbl:snrate}}
\tablehead{
\colhead{Reference} & \colhead{Redshift} & \colhead{Rate (SNu)}}
\startdata
Cappellaro et al. (1999) & 0     & $0.18\pm0.05$ \\
Dilday et al. (2008)\tablenotemark{a}     & 0.09  & $0.246^{+0.076}_{-0.060}$ \\
Madgwick et al. (2003)   & 0.098 & $0.196\pm0.098$ \\
Blanc et al. (2004)      & 0.13  & $0.125^{+0.044+0.028}_{-0.034-0.028}$ \\
Hardin et al. (2000)     & 0.14  & $0.22^{+0.17+0.06}_{-0.10-0.03}$ \\
{\bf This work}                & 0.20  & $0.257\pm0.028\pm0.024$ \\
Botticella et al. (2008) & 0.3   & $0.22^{+0.10+0.16}_{-0.08-0.14}$ \\
\enddata
\tablenotetext{a}{$B$-band luminosity density of $j_B=1.19\times10^8\;L_\sun^B\;{\rm Mpc}^{-3}$ was used to convert from the value per comoving volume unit.}
\end{deluxetable}

\begin{deluxetable}{cccccccccc}
\tabletypesize{\scriptsize}
\rotate
\tablecaption{Predicted SNe Ia rate in the two component model\label{tbl:rate}}
\tablewidth{0pc}
\tablehead{
\colhead{} & \colhead{} & \colhead{} & \colhead{} & \multicolumn{3}{c}{Mannucci et al. model} & \multicolumn{3}{c}{Sullivan et al. model} \\
\colhead{morph. type} & \colhead{$g-r$} & \colhead{$j_r$} & \colhead{$L_K/L_r$} & \colhead{$N_A$ (delayed)} & \colhead{$N_B$ (prompt)} & \colhead{$N_A+N_B$} & \colhead{$N_A$ (delayed)} & \colhead{$N_B$ (prompt)} & \colhead{$N_A+N_B$}\\
\colhead{} & \colhead{} & \colhead{$10^8L_r(\sun){\rm Mpc}^{-3}$} & \colhead{} & \colhead{$10^{-5}{\rm Mpc}^{-3}{\rm yr}^{-1}$} & \colhead{$10^{-5}{\rm Mpc}^{-3}{\rm yr}^{-1}$} & \colhead{$10^{-5}{\rm Mpc}^{-3}{\rm yr}^{-1}$} & \colhead{$10^{-5}{\rm Mpc}^{-3}{\rm yr}^{-1}$} & \colhead{$10^{-5}{\rm Mpc}^{-3}{\rm yr}^{-1}$} & \colhead{$10^{-5}{\rm Mpc}^{-3}{\rm yr}^{-1}$}
}
\startdata
E,S0    & 0.75 & 0.62 & 3.32 & 0.720 & 0.000 & 0.720 & 0.882 & 0.052 & 0.935 \\
S0/a-Sb & 0.64 & 1.00 & 2.77 & 0.970 & 0.304 & 1.274 & 1.052 & 0.189 & 1.240 \\
Sbc-Sd  & 0.51 & 0.37 & 2.52 & 0.326 & 0.495 & 0.821 & 0.326 & 0.088 & 0.414 \\
Im      & 0.36 & 0.02 & 2.15 & 0.015 & 0.127 & 0.142 & 0.013 & 0.006 & 0.019 \\
\hline
Total   &      & 2.01 &      & 2.031 & 0.926 & 2.957 & 2.273 & 0.335 & 2.608 \\
\enddata
\end{deluxetable}

\clearpage

\clearpage

\end{document}